\documentclass[12pt]{aa}
\usepackage[english]{babel}
\usepackage{psfig,graphicx}

\onecolumn

\begin{document}

\title{Spectroscopy of 13 high mass stars in the Cyg~OB2 association}

\author{E.L.\,Chentsov, V.G.\,Klochkova, V.E.\,Panchuk, M.V.\,Yushkin}

\institute{Special Astrophysical Observatory RAS, Nizhnij Arkhyz,  Russia}

\date{\today}	     

\abstract{Aiming to explore  weak spectral features of stellar and interstellar origin we used 
the NES  echelle spectrograph  of the 6--m telescope to obtain  high--resolution spectra for 
13 hot  O3--B4 stars in  the Cyg\,OB2 association, including a high luminous star No.\,12. 
Velocity fields in the atmospheres and interstellar medium,  characteristics  of optical 
spectra and line profiles are investigated. The cascade star formation scheme  for the association 
is confirmed. Evidence is presented suggesting that the hypergiant Cyg~OB2 No.\,12 is an LBV 
object and that its anomalous reddening has  a circumstellar nature.}

\authorrunning{Chentsov et al.}
\titlerunning{Spectroscopy of 13 high--mass stars in the Cyg~OB2 association}

\maketitle

\section{Introduction} 

Among hypergiants of the Galaxy, those belonging to stellar associations  are most attractive for study. 
Membership in a stellar group makes it easier to estimate a star’s age, luminosity, and center of mass 
radial  velocity, which is needed to introduce a zero point  for the system of velocities observed in 
the stellar  atmosphere and wind.
In addition, studies of stars with extreme luminosities and mass loss rates can naturally be expanded
to their influence on the association’s gas and dust component and the star formation process. In the
present paper, we address this two-sided problem for the enigmatic star No.\,12 and other members of the
Cyg~OB2 association. This association (initially named VI~Cyg; $\alpha_{2000}=20{\rm ^h} 33.2{\rm ^m}$, 
$\delta_{2000}=+41.32{\rm ^o}$,  l\,=\,80.2${\rm ^o}$, b\,=\,+0.8${\rm ^o}$) was discovered 
in the middle of the 20-th century and still remains one of the most actively studied
objects in our Galaxy. Refinement of its principal characteristics (size, population, age, etc.) is still
underway; the list of frequently cited publications given below shows that this refinement has proceeded
especially intensively in recent years.

\subsection{History of stellar population studies for Cyg~OB2}  

1953: M\"unch and Morgan [1] noticed a concentration of OB stars to the northwest of $\gamma$\,Cyg on a photograph taken 
through an objective prism. A year later, it acquired the status of an O--association: it turned out that 13 of the first 
14 stars identified in it were O stars [2]. The considerable extinction towards
Cyg~OB2 ($4.5^m <$\,Av\,$< 6.8^m$) is not high enough to severely impede high resolution spectroscopy for the
brightest association members, and the accompanying reddening even makes it easier to use a low dispersion 
objective prism or three color photometry to search for fainter hot stars [3, 4].

1966: Based on their photographic UBV photometry and counts of faint stars on Palomar Sky Survey
images, Reddish et al. [5] increased the projected area of Cyg~OB2 on the plane of the sky to a third
of a square degree (the linear diameter was about 20\,pc); the number of possible OB members of the
association increased to 300.

1991: Although Massey and Thompson [6] restricted their study to the central part of the association, their more 
accurate CCD photometry enabled them to reliably identify its 108 brightest members, for 76 of which they performed 
2D spectral classiﬁcations based on moderate-resolution spectra. This
study made it possible to accurately trace the main sequence (MS) in the Hertzsprung–-Russell (HR)
diagram, and to estimate the limiting mass for an association member to be 85${\mathcal M_{\sun}}$  and the 

2000: Suspecting that Reddish et al. [5] had studied only a relatively transparent  ``window'',
Knodlseder [7] attempted to broaden it using counts in the near-IR from the 2MASS survey. According 
to his estimates,  the diameter of the association is about 60\,pc, the number of OB members is about 2500,
and the total mass is from 40 to 100 thousand ${\mathcal M_{\sun}}$, so that it can be 
considered the first young globular cluster detected in our Galaxy.

2002: Comeron et al. [8] added IR photometry with low-resolution IR spectroscopy. They confirmed
Knodlseder’s estimates, but drew more cautious conclusions, reporting Cyg OB2 to be only one of the
Galaxy’s highest-mass clusters, ``perhaps comparable to a young globular cluster.''

2003: Hanson [9] expressed stronger doubts about the unique status of Cyg OB2, called a super-star
cluster. She was able to provide MK classiﬁcations for the
brightest OB--star candidates, based on the shortwavelength part of the spectrum; these are indeed
early  type stars, but most are appreciably older than the main population of the association and are located
in its extended halo.

2007: The spectroscopic survey of Kiminki et al. [10] doubled the number of OB stars with MK classiﬁcations 
and identified as members of Cyg~OB2. This survey also made it possible to use radial velocities
as an additional criterion of association membership: the velocity dispersion was found to be moderate
(2.4\,km/s), as is typical of open clusters.

2008: Photometry in the visible and near-IR and low-resolution spectroscopy revealed two types of
stellar condensations, both spatially close to the association and probably related to it evolutionarily, 
although they are located in its peripheral regions. The first contains A0--A5 stars, which are several million
years older than the high mass Cyg~OB2 stars [11], while the second contains late type stars that have
not yet reached the MS and were presumably formed under the influence of the hot, higher-mass stars of
the association [12]. Comeron et al. [13] repeated their photometric and spectroscopic study in an annular 
zone between $1^{\rm o}$ and $2^{\rm o}$  from the association center and found evolved BIII, Be, and WR stars
there, in addition to field stars; in their opinion, this ``conclusively dismisses the case for a large extent
of Cygnus~OB2 beyond the boundaries of its central concentration.'' However, Negueruela et al. [14] believe 
that star formation has proceeded in small bouts over an extended period of time in the central part of
the association as well: most of the stars have ages of about 2.5~million years; stars that formed 5--7 million
years ago are also present, as well as quite young stars (or rejuvinated, blue stragglers?).

2010: Wright et al. [15] augmented the list of Cyg~OB2 members with new young stars, identified
from their X-ray radiation, and estimated the total mass of the association to be 30 thousand ${\mathcal M}_{\sun}$. 
They found new evidence that the current O stars were produced in the latest phase of triggered, rather than
burst, star formation. Thus, Cyg OB2 is a rich association but not a super-cluster, as is natural for an 
aggregate formed in a disk galaxy far from its center [16]. In a supercluster, stars with a total mass of 
at least 100 thousand  ${\mathcal M}_{\sun}$ are concentrated in a volume not exceeding several cubic parsecs. 
The object closest to us that approaches a supercluster according these parameters is Westerlund~1~[17]. 
Its spatial structure differs drastically from that of Cyg~OB2: a direct photo (in red light: 
Av$\approx 10^{\rm m}$~!) shows a very compact group of stars, while two small subclusters of the Orion 
Trapezium type were identified in the central region of Cyg~OB2 only recently [18], and, taken together, 
they contain no more than one-third of all O stars of the association.

\subsection{The problem of star No.\,12}

Among the brightest members of Cyg~OB2, star No.\,12 (using the numbering of Schulte [3]) has
attracted the most attention. Immediately after the discovery of the association, Morgan et al. [19] noted
this star’s striking combination of a red color (it is considerably redder than its neighbors, stars No.\,5
and No.\,9), early spectral type, and very high luminosity. It was soon found that Cyg~OB2 No.\,12
(hereafter called star No.\,12 for brevity) also stood out among the association’s stars in its abnormally
high degree of polarization [20]. Having moved to the near-infrared for spectroscopic classiﬁcation of
star No.\,12 and selected several comparison stars, Sharpless [21] obtained the spectral type B5, absolute
magnitude Mv$\approx -9.5^{\rm m}$, and interstellar extinction Av\,$\approx 10.1^{\rm m}$. 
No significant refinements of these estimates have since been found necessary. If there
were no absorbing matter in the line of sight, this star would rival Sadr and Deneb in brightness.
In the HR--diagram, star No.\,12 is above the Humphreys--Davidson limit and is one of the brightest stars 
in the visible range in our Galaxy, making it natural to suspect that it is a luminous blue variable (LBV). 
Numerous attempts to confirm the LBV status for this star have not brought a ﬁnal result. In 1984, Leitherer et al. 
[22] noted that the wind from star No.\,12 was similar those from S~Dor variables.
According to Massey and Thompson~[6], it is an incipient LBV star. Humphreys and Davidson~[23]
believe star not to be a full-ﬂedged LBV star. Van Genderen [24] calls it a weak-active star. The result is that star 
No.\,12 was still listed as only a candidate  LBV object in the review of Clark et al.~25] in 2005.

Of the principal characteristics of LBVs, the star has a high luminosity and a high mass-loss rate, an early
spectral type, and brightness variability. Variability has been detected both in the blue [26] and in the
red [27], with mean magnitudes of 14.7$^{\rm m}$ and 9.0$^{\rm m}$, respectively, and brightness variations 
reaching 0.3$^{\rm m}$.
However, to our knowledge, it remains unknown if the star’s brightening is accompanied by reddening -- an
effect noted even for the lowest-amplitude LBV objects [28]. 

The first information obtained about the wind from star No.\,12 was contradictory. Judging from the radio 
emission from free--free transitions discovered in 1980 by Wendker and Altenhoff~[29], it is as strong as 
the wind of P~Cygni [30], but is not manifest via an IR excess for some reason. This contradiction was removed 
in~[22] using a cool-wind model (Te\,=\,5000\,K for Teff\,=\,13600\,K), however, with an unnaturally high limiting 
outflow velocity, $\approx 1400$\,km/s. 
This velocity is an order of magnitude lower for LBV objects during their maximum visual brightness; their winds
are not only cool, but also slow. The above estimate is probably erroneous: it is based on low-quality spectra
of star No.\,12 where a depression near 6532\,\AA{} was taken to be a blueshifted H$\alpha$ component~[31, 32]. 
This depression could result from a local  defect and a group of telluric H$_2$O absorption festures 
overlapping the diffuse interstellar band (DIB) at 6534\,\AA{}~[33]. 
Recent high resolution spectra have no anomalies near 6532\,\AA{},  and the absorption H$\alpha$ 
components indicate a limiting velocity $\approx 150$\,km/s~[34].

High angular resolution centimeter wavelength radio observations reveal variability and asymmetry of
the wind from star No.\,12; at 6 cm, the source is elongated North--South ($0.12\times 0.24 ^{''}$), 
and a condensation can even  be noted in its southern part [35]. 

The question of star No.\,12 is an LBV star may be related to its anomalous reddening. 
The star’s color excess E\,(B--V) is at least 1$^{\rm m}$ larger than those
of other stars of the association. The hypothesis that this additional reddening has a circumstellar origin, 
which would provide a strong argument for star No.\,12 being an LBV object, has not yet found reliable
observational confirmation. The star has been identified with the IR--source IRAS\,20308 + 4104 and the
presence of cool dust in its envelope is suspected [36], but direct images of the dust shell have not yet been
detected in IR frames. 

This paper describes our own high resolution spectra for the 13 brightest stars in the region of the Cyg~OB2 
association. These spectra  will be used for model analyses of the atmospheres and winds of particular stars, 
as well  as the derivation of empirical Vr\,(r)  relations and their zero points, i.e., the radial velocities of the 
stellar centers of mass (Vsys). These relations can potentially help us approach an understanding of the nature 
of star No.\,12, where the problem of Vsys is especially acute~[34]. Unfortunately, our spectra do not contain  
emission lines from HII regions, but they are rich in absorption lines that provide information on the structure 
and kinematics of the interstellar medium in the direction of the association and, in the volume of the 
association proper, information on the interaction of stellar winds with the ambient matter.

\subsection{Observations, reduction and analyses of the spectra}

Table~1 presents information on the available spectroscopic material. The first five columns contain
the star numbers from Schulte~[4], dates when the spectra were taken, names of spectrographs used,
spectral resolution, and spectral range registered. The last column contains references to publications where
fragments of low-- or moderate--resolution spectra for corresponding stars can be found. The vast majority
of our spectra were taken with the NES echelle spectrograph [42] at the Nasmyth focus of the 6-m telescope at
the Special Astrophysical Observatory. These observations used a 2048\,x\,2048 CCD chip and image
slicer~[43]. The spectroscopic resolution was $\lambda/\Delta\lambda\ge 60000$ and the signal-to-noise ratio 
was S/N$\ge$100.
In addition to these spectra, we also used three spectra taken with the MAESTRO Coud\`e echelle spectrograph of the 
Mt.~Terskol Observatory [43] and a spectrum taken with the PFES echelle spectrograph [44] at the primary focus of 
the 6--m telescope.

We extracted the one--dimensional spectra from the two--dimensional echelle frames using the modified~[45] 
ECHELLE  context of the MIDAS software package. We removed cosmic rays via median averaging of pairs of 
successive  spectra. The final reduction (continuum tracing and continuum normalization,
measurements of the radial velocities and equivalent widths) was done using the DECH20 software package [46], 
traditionally used at the SAO. In particular, this package includes a procedure for applying mutual wavelength 
shifts of the direct and mirror profile images, enabling measurement of the radial velocity
of any feature of the profile.
We performed the wavelength calibration using spectra of a Th--Ar hollow cathode lamp. Control and
corrections of the position shifts between the stellar and lamp spectra were performed using 
telluric absorption and emission features; the residual systematic errors of the resulting radial velocities are no more 
than 1\,km/s. All velocities discussed in this paper are heliocentric. The procedure used to measure
the radial velocities Vr from the NES spectrograph spectra and sources of uncertainties are discussed in
more detail in~[47].

Since the question of correct laboratory wavelengths remains important for spectroscopy of O stars
(values for high-order C, N, and O ions in various recent publications differ by as much as several km/s), 
thus we used the wavelengths from the NIST database 
($www.nist.gov/physlab/data/asd.cfm$), after checking them using the spectra of the narrow line 
standard stars 10~Lac (O9 V) and HD~163800 (O7\,III). The spectrum of the first star was taken from 
the NES archive and that of the second star from the library of the UVES spectrograph~[48]. 
We used for the DIBs the wavelengths of [49], considered to be most reliable. 
Fragments of our spectra are displayed in Figs.\,1 and 2. The continuum intensity in all the figures 
is taken to be 100. Selected data on the stars used to estimate their distances using the standard 
photometric parallax technique, along with  the distances themselves, are presented in Table~2. 

The spectral types for the O3--O8 stars were taken  from the references given in the last column of Table~1; 
for the remaining stars, we refined these using  available spectra. This refinement was motivated by doubts 
concerning association membership for stars No.\,2, 10, and 18: they are located in its outskirts of
the associaton, and their points in the HR--diagram are to the right of the MS but belower the upper
MS. Differences between the new spectral types and those earlier used are not large, but quite distinct.
For example, stars No.\,2 and No.\,18 have the same type, B1\,I, in the spectroscopic survey of Kiminki et
al. [10], while the types in our study are B0.5\,II and B1\,Ib, respectively. Figure~1 shows that the N\,II
and Al\,III absorption becomes stronger while the C\,IV absorption disappears in the latter compared to
the former star. Figure~2 illustrates the refinement of the luminosity class. Stars No.\,10 and  21
are considered to be supergiants in [9], but their H$\alpha$ profiles indicate that star No.\,10 is indeed a
supergiant, but star No.\,21 is a dwarf.

However, our new classiﬁcation of these three peripheral stars resulted in only small displacements of
their data points in the HR--diagram, thus preserving their assumed evolutionary status of older-generation
stars in the Cyg~OB2 association~[14]. Their estimated distances were likewise not significantly influenced; 
Table~2 shows that these distances are the same as the distances to the association’s central
stars, within the errors; i.e., at least the interstellar components of their spectra should be no less 
informative than for the other stars.

\section{Main results}

\subsection{Stellar radial velocities}

The radial velocity data  Vr are collected in Table~3 (values are not presented for Cyg~OB2 No.\,12, as we
will discuss them separately). The uncertainties of the Vr values for individual lines and averages over
groups of lines can be judged from the scatter and number of symbols in Fig.\,3, which displays Vr(r)
relations for several stars. Along with the S/N, the accuracy of the Vr measurements is determined by a
line’s width and intensity; because of this, the second column of Table~3 presents our estimates of the 
halfwidths of weak lines at the continuum level. For five  stars also considered in [37], these values coincide
with the V\,{\it sin\,i}  estimates from [37] within the errors. Our O--star spectra provide 6--12 measurable 
lines, and our B--star spectra 30--50. Most of these are weak (deviations from the continuum level
of no more than 20\%). 
In all cases when the intensity difference in a line considerably exceeded the noise
level, we measured the velocity for the absorption core or emission peak. This is especially important for
early O stars, in which the profiles of sufficiently deep absorption lines are distorted by their winds. 
Figure~4 shows that these profiles are asymmetric: the red slope of the profile is steeper than the blue one, 
the red wing is shortened, and the blue wing is extended; i.e., there is a positive velocity gradient with depth. 
These signatures of direct P~Cygni profile are the same for the H\,I, He\,I, and even He\,II lines. 
Velocities for lines with such profiles are labeled ``P'' in Table~3.

All the velocities presented in Table~3, with the exception of those for interstellar NaI\,(1) absorption
features, are rounded to integer km/s. The remaining columns contain  mean values derived from the C\,III, 
IV and N\,IV emission lines (column~3); the limiting mean velocities from absorption lines, approached in
the transition from the deepest to the most shallow lines (r$\approx$100) (column~4), derived using plots 
of the Vr\,(r) relations;  the velocities derived from He\,I lines, from line  ,
the single He\,I\,(11) 5875.72\,\AA{}  for stars No.\,7, 22,  4, and 10 and averaged from several
lines for the other stars (column~5); the velocities for line cores or the absorption components of P\,Cygni
profiles for the H$\beta$ and H$\alpha$ lines (column~6 and 7, respectively);  
for comparison, the velocities of the same stars from other papers, with corresponding references (columns 11--12). 
For the two--spectra spectroscopic binaries Cyg~OB2 No.\,5  and No.\,8, these are given with phases taken into
account.

The differential line shifts we detected are no greater than 10\,km/s, with the exception of two cases
with considerable Balmer progressions (stars No.\,9 and No.\,10). However, only the data from column~4 or
the weighted mean values of the data from columns 4 and 3 were used as ``systemic'' velocities, 
Vsys, describing the star as a whole. The mean Vsys for seven reliable association members is 
$-10.7 \pm 0.7$\,km/s.
This differs considerably from the mean value for the velocities found for the same stars in [10] 
($-15\pm 1.8$\,km/s), but coincides, within the errors, with the Vsys value found in~[10] for several 
dozen stars ($-10.3 \pm 0.3$\,km/s).

\subsection{Radial velocities for the interstellar medium}

Lines of the interstellar medium are mainly represented in the NES spectrograph spectra by absorption 
lines of the Na\,I\,(1) doublet and numerous DIBs. In addition, the spectrum of star No.\,5 acquired on
September 14, 2011 exhibits lines of CH and CH$^+$, and other spectra of the same star also contain K\,I
and Ca\,II lines. Interstellar emission lines are not accessible to measurement.
Revealing weak features of Na\,I\,(1) lines and measuring their parameters requires careful account
of the telluric component of the spectrum. It is easy to recognize and remove the  atmospheric Na\,I
emissions that sometimes distort the lowest parts of the profiles. To remove water--vapor absorption lines,
we took spectra of several suitable comparison stars with the same NES spectrograph and reduced them
using the same techniques as for the Cyg~OB2 stars.
One of these is HR\,4687, whose spectrum contains broad stellar ($V \sin i = 243$\,km/s [53]) and very weak
interstellar lines, so that the spectrum in the range of the Na\,I\,(1) doublet exhibits an undistorted set of
atmospheric water--vapor absorption lines. Figure~5 displays a fragment of this spectrum, overlaid with the
corresponding spectrum fragment for star No.\,6 from Cyg~OB2.

Profiles of the stronger D2 line of the Na\,I\,(1) doublet are shown in Fig.\,6 for stars No.\,9 and No.\,22;
Table~4 presents the parameters of these profiles for all the program stars. Figure~6 shows that, in each
of the profiles, it is possible to identify a saturated trapezium--shaped component with shallower triangle
components on either side. Figure~7 displays profiles of the weaker D1 line for 4 other association members. 
The residual intensity r is still present in its core: $1 < r < 4$ in the range $-13< Vr < 0$\,km/s.
Column~8 in Table~3 contains radial velocities for the cores of the main Na\,I\,(1) components taken as a whole 
(i.e., for their bisectors). However, since these components are fairly broad (15--22 km/s), and their
slopes are steep and sometimes almost vertical, we found it useful to supplement these Vr values with
those for the blue and red slopes (V$_{\rm blue}$ and V$_{\rm red}$). These are presented, one below the other, 
in the third column of Table~4 for the r\,=\,10 level in the D2 profiles of the association stars and 
for the r\,=\,70 level in a foreground star. The following columns of Table~4 present the radial velocities
and central depths of the side D2 components. Note that the side components are much weaker than the main ones; 
they are not visible in the well-traced profiles of the KI(1) lines presented in [54], and we were not able to 
detect them in our own spectra, even for the deepest and narrowest DIBs.

We can identify several indications that the deep, broad components of the Na\,I\,(1) lines are blends of
narrower components. Multicomponent profiles of unsaturated interstellar lines (K\,I, etc.) are known
for some of the stars  (the corresponding references can be found in the last column of the line
in Table~4). For example, the profile of the K\,I~7699\,\AA{}  of star No.\,12 was resolved in [54] into 5 
individual components with velocities from $−14$ to 9\,km/s. The same is true for Na\,I\,(1) profiles if 
we consider bright stars located in the vicinity of the association in projection on the sky, 
but less distant than it. The main profile components of the stars P\,Cyg and 55\,Cyg,
whose distances are close to those of the Cyg~OB2 association, are as saturated and broad as for the stars
of our sample; for the less distant stars $\alpha$\,Cyg and $\gamma$\,Cyg, they are resolved into shallower 
and narrower (down to 1--2 km/s) components~[57]. 

Finally, our spectra also provide some evidence that the main components have structure. Our spectroscopic 
resolution and exposure times were insufficient to detect features in deep cores, but do display an increase of
the main components’ width due to the nearest side components. The data in Table~4 demonstrate that
the largest depth (R\,$\approx$\,70) is reached by the components closest to the main ones, and Fig.\,7 shows how
these merge with the latter. This is obvious for the components with Vr\,$\approx -28$\,km/s in the spectrum of
star No.\,21 (right panel of Fig.\,7) and is not excluded for the red profile slopes in the spectrum of star 
No.\,6, which are displaced by 6\,km/s relative to those for star No.\,16 (the left panel of Fig.\,7).

Moreover, we also have evidence from the relative intensity of the main components of the Na\,I\,(1)
absorptions: the shorter wavelength components among these are usually deeper than longer wavelength ones. 
Indirect evidence for this is provided by the K\,I~(1) profiles in the spectra of stars No.\,5
and 12~[54] and the evolution of the Na\,I\,(1) profiles with distance in the direction of Cyg~OB2,
which can be followed in high resolution spectra for bright stars. The nearby stars $\delta$\,Cyg 
(d\,$\approx$50\,pc) and 57\,Cyg (d\,$\approx$\,150\,pc) have single component, narrow, 
unsaturated D2 profiles, and the heliocentric Vr radial velocities derived from them are 
Vr\,=\,$-18$ and $-16$\,km/s, respectively. For the more distant stars 59\,Cyg (d$\approx$0.4 kpc), 
$\alpha$\,Cyg, and $\gamma$\,Cyg (d\,$\approx0.5$\,kpc for both), in addition to isolated 
components at Vr\,$\approx -21$\,km/s, close groups of components between Vr\,$\approx -12$\,km/s and 
Vr$\approx 2$\,km/s  appear, with the depths of  decreasing with wavelength~[58].
The cores of the D1 lines of most of our program stars also show an increase in the residual intensity,
from r$\approx 2$ at Vr$\approx  -11$\,km/s to r\,$\approx 4$ at  Vr\,$\approx -2$\,km/s. 
The profiles of the weaker CH$^+4232$\,\AA{} interstellar lines (r\,=\,71 and 60,
and CH~4300\,\AA{}, respectively) in the spectrum of star No.\,5 obtained on September~14, 2011 are clearly 
asymmetric:  the red slopes rise from the cores at Vr\,$\approx -12$\,km/s much less steeply than the blue slopes.

This can be related to the fact that the widths of the main components differ appreciably in our spectra
(by a factor up to 1.5 for stars No.\,2 and 6).
Table~4 (column~3) shows that the D2 line at r\,=\,10 is the narrowest for star No.\,2:  $-17.0 < Vr <-1.8$\,km/s 
(Fig.~6), and there is no V$_{\rm blue}$ or V$_{\rm red}$ value for any other star in this interval. This suggests we
can use the above velocities as zero points, and the shifts of the blue and red profile slopes relative to these
velocities as a measure of the profile width. Figure~8 displays the relative values of these widths for stars
in the central region of the association. The sizes of the filled circles and squares in this figure are proportional 
to ($-17-$V$_{\rm blue}$) and (1.8\,+\,V$_{\rm red}$), respectively. The squares are larger than circles for all our stars: the
main profile components are more broadened toward the red than toward the blue. The same is true
for the differences from star to star: the largest and smallest amounts of broadening toward the red differ
by 4.9\,km/s, and toward the blue by only 2.6\,km/s. The difference is still stronger for the D1 line: 
6.3 and 2.5\,km/s, respectively. 

If the profile of the Na\,I\,(1) interstellar line is formed from an almost continuous series of
narrow components, why should we single out their central  condensation and consider it separately 
from the side components? Apart from the obvious methodical  advantage of the clearer outlines 
of the profiles, the central component also stands out because it is formed along the entire 
line of sight, in contrast to the side components, which appear at the more distant 
end of the line of sight, possibly only close to Cyg~OB2. This last suggestion is based on a
comparison of spectra presented in [57, 58] and taken from the SAO spectroscopic archive: we can see
no side components for stars in the direction of the association located at distances up to 1\,kpc, 
whereas they are present for stars that are spatially close to the association. In particular, the Na\,I\,(1) 
profiles of the supergiants HD\,194839 and HD\,194279, which are separated from the center of Cyg~OB2 
by $1^{\rm o}$ and $2^{\rm o}$, respectively, and lie at a distance of 1.8\,kpc from the Sun, have 
components with Vr\,$\approx -33$\,km/s   that are comparatively deep as for stars No.\,9 and No.\,22, 
as well as weaker components with Vr\,$\approx -55$\,km/s, and the star HD\,195592 (at a distance of 
1.8\,kpc and separated from Cyg~OB2 by $3^{\rm o}$) has a component with Vr\,$\approx 13.5$\,km/s.

Thus, it is natural to compare the intensities of main components of the Na\,I\,(1) profiles in the spectra
of our program stars to the amount of interstellar absorption and the intensities of DIBs, which also 
accumulate along  the entire line of sight. Recall, however, that the latter are not accumulated uniformly with
distance. This is demonstrated for the interstellar extinction Av, for example, by Fig.~23 in [59]: the
slow increase of Av to d\,$\approx$\,1.5\,kpc in the direction of Cyg~OB2 is followed by a jump 
by 3$^{\rm m}$ in  a layer only about 100\,pc deep (an ``extinction wall''). This nonuniformity is also 
characteristic of  the variations of the depths and equivalent widths of the DIBs: half of these quantities 
accumulates  for the Cyg~OB2 stars by d\,$\ge$\,1.5\,kpc. 

Table~5 presents the equivalent widths and depths for the Na\,(1)\,(1) doublet and for several 
DIBs (seven of more than 100 identified in the entire usable wavelength range). Figure~9 relates 
the intensities of Na\,I\,(1) lines and their main components to the depth of one of the strongest 
interstellar bands, the DIB. 
Since the equivalent width, which is a 5797\,\AA{} natural intensity measure for a line as a whole, loses
its definiteness for the main component, we replaced this with the profile width at the r\,=\,10 level; for
better accuracy, we used sums for the D1 and D2 lines. The DIB in the central part of the depth of 
the 5797\,\AA{}  association displayed in Fig.~9 varies by a factor of almost two, and the growth with 
its increase is obvious, as expected, both for W and for $\Delta$\,Vr. 

However, stars with similar DIB  depths are not randomly distributed across the association field, 
and instead form compact groups, delineated  by rectangles in Fig.~9. Only 2 to 4 DIB intensities 
are available in each of these groups, but it is possible to use many more Av estimates. 
Table~2 in [10] presents from 15 to 20 Av values for fields  3$^{\rm '}$--5$^{\rm '}$ in size  
that include the selected  groups, and Fig. 4 from [10] demonstrates that  these values 
are close to each other within the fields, but, like R(5797), differ considerably in different 
fields. The first of these, with stars No.\,4, 6, 16, and 21, is in a relatively transparent field 
that is clearly seen in interstellar-extinction maps [60, 61]; the second field, with stars No.\,7 and 8, 
and especially the third field, with stars No.\,9, 22, and  18, are closer to its southern boundary, and their
mean Av values are 4.3$^{\rm m}\pm 0.3^{\rm m}$,  5.2$^{\rm m}\pm 0.2^{\rm m}$, and 
6.8$^{\rm m}\pm 0.2^{\rm m}$,  respectively. As follows from Fig.\,9, as well as Figs.~7 and 8, 
the first group exhibits a large scatter of Na\,I\,(1) profile widths, i.e., of intensities of the main 
components at r\,=\,10. This scatter is much lower in the second group and close to the
measurement uncertainties in the third group. It is not surprising that the $\Delta$Vr values become 
closer for the two latter groups: both are clusters identified within the association, while an increased 
star density in the first group was noted only near star No.\,4~[14, 18]. However, if we return to Table~4, 
we ﬁnd  that the scatter of V$_{\rm blue}$  is close to the measurement uncertainties for all three groups,
and  the mean values for the first and third groups differ by only 0.8\,km/s ($-$18.0 and $-$18.8\,km/s, 
respectively).

In contrast to the widths of the lower profile sections, the total equivalent widths do not become
closer in the transition from the first to the third group (Fig.\,9). This already indicates that the 
equivalent width differences are due to the presence of side components in the profile. Indeed, the equivalent
widths of the Na\,I\,(1) lines are correlated with the sum of the depths of their side components, and
Figs.~6 and 7 provide a direct confirmation of this: the contribution of the side components to the equivalent
widths is larger for star No.\,9 than for star No.\,22, as is also true for star No.\,6 compared to star No.\,16 
and for No.\,21 compared to star No.\,4.

The relation between intensity and position is clearly shown by the side components of the Na\,I\,(1)
line profiles: the depth of the side components decreases with distance from the main component, 
toward both the blue and the red. At the same time, at least the blue components are positioned similarly to
the main components, as is shown by Fig.\,10, which presents the dependence of the component depths
and radial velocities on the star’s position in the central region of the association. To demonstrate this
positional segregation more clearly, only sufficiently
well separated components with distinct minima (like those for star No.\,9 in Fig.\,6) were selected for this
figure; with our resolution, these appear at V$_{blue}-$Vr, distances exceeding 10\,km/s. The zones of the
component mean values Vr\,$\approx -50$ and $-32$\,km/s are clearly separated: the former components are weak
and observed only in the above relatively transparent window (upper left part of Fig.\,10); the latter are
much stronger, and are observed only towards the boundary, where the transparency is poorer (lower
right part of the figure). It follows from Table~4 that the velocities and depths of the red side components
are close only for stars No.\,4, 6, 16, and  21 (Vr\,$\approx 19$\,km/s, R\,$\approx 20 -30$); 
the velocities of the rest of the stars are scattered from 12 to 41\,km/s, and their depths range from 
9 to 66.  The total velocity range determined with the side components of the Na\,I\,(1) lines is 
at least 80\,km/s ($\pm 40$\,km/s relative to the mean radial velocity of the association stars). 

Our data naturally supplement the picture of the expansion of the interstellar medium and induced star 
formation under the influence of the winds from the brightest association stars, presented by molecular 
radio~[62] and H$\alpha$ spectroscopy~[63]. Judging from the relation between the shifts and depths of 
blue components, the outflow velocity of the gas is the highest where its density is lowest. 
The red components are unrelated to the expansion, in contrast to the H$\alpha$ emission profiles, 
and can be related to the formation of new stars and clusters of new stars. In particular, is it due 
to chance that star No.\,9, which is projected on a dust cloud and molecular
clump (``clump 1'' in~[64]), stands out in terms of the intensity of its red components?

\subsection{Star No.\,12}

Strong H$\alpha$, weak H$\beta$, and two weak Fe\,II emission lines near 7500\,\AA{} are present in our spectra 
of the Cyg~OB2 star No.\,12. There are only 2--3 strong He\,I lines (R\,$\approx$\,30) and the CII\,(2) and 
SiII\,(2) doublet (R\,$\approx$\,20) absorption lines; the depths of 90\,\% of the remaining lines do not 
exceed 10, and do not exceed 4 for half of them. As for the hotter stars in the association, the profiles of the strongest absorption
lines of star No.\,12 are asymmetric, with the same kind of asymmetry: the red slope is steeper than the
blue one (Fig.~11). This suggests that their formation zone is not limited to the photosphere, and includes
the base of the wind. Having in mind this possibility, we used our additional spectroscopic material
together with spectra from [65] for the range 4100--4800\,\AA{}, which is not included in our observations,
to estimate the spectral type of star No.\,12, based solely on the weak, photospheric absorption lines. We
compared the line groups N\,II and S\,II, Fe\,III and Fe\,II,~etc., whose intensities reach their maxima in
earlier or later spectral subtypes relative to the one we determined. This indicated somewhat earlier subtype
than we found in [34]: B4 instead of B5. This result coincides with the estimate of Clark et al.~[65]. This
is also confirmed by a comparison of the spectrum of star No.\,12 with the spectra of HD\,168625 (B5.5\,Ia0 
[66]) and  HD\,80077 (B2.5\,Ia\,0 [67]), eliminating the effect of luminosity: both stars are hypergiants, 
the former being cooler  and the latter hotter than star No.\,12.

Velocities vary from epoch to epoch and also differ  between groups of lines. Figure~12 displays the Vr\,(r)
relations for the two dates with the largest radial velocity deviations: Vr\,(SiII)\,=\,$-29$\,km/s 
(April~12,~2003)  and Vr\,(HeI)\,=\,0\,km/s (September~26,~2010).  

The velocity time variations in our data range from 13 to 21\,km/s. In all cases, V\,(r$\rightarrow 100)$ is 
below Vsys, which can be interpreted as expansion with variable velocity,  even of the deep atmospheric 
layers where the weakest absorption lines are formed. This seems to be in contradiction with 
the shifts of the strong relative to the weak absorption features: these shifts are lowest for the SiII~(2) doublet, 
which even has a negative mean shift ($-1.4$\,km/s), but positive (+4.8\,km/s) for the CII~(2) doublet, increasing
to 13\,km/s for the strongest HeI lines. However, as was mentioned above, these numbers were obtained from profile 
cores  carrying information about the outer atmospheric layers, while their upper parts also indicate expansion. 
Figure~11  shows that the wings are displaced to the blue relative to the cores, and are stronger for deeper lines. 
As they rise toward the continuum, the profile bisectors deviate toward the left, to the velocities derived 
from the weakest lines or even lower. The blue wing is extended and the red one shortened (a hint of 
a P~Cygni profile). For the strongest of the absorption features, the HeI\,5876 line, we find 
Vr\,$\approx -150$ and 60\,km/s for the blue and red wings at the r\,=\,98 level in the spectrum taken 
on September~15,~2011, when the differential line shifts were the smallest; for comparison, the NII\,5679 
and SII\,5433 lines had Vr\,$\approx -70$ and  +42\,km/s at the same residual intensity level.

The strong absorption features vary with time in their positions, intensities, and even profile shapes.
Sometimes, apparently at the times of the greatest deepening of the core, the derived velocity approaches
the velocity of the dip in the upper part of the H$\alpha$ line.
One such case, already noted in [34], was detected line  on April~12, 2003 for the Si\,II~6347\,\AA{} 
(Fig.\,11). Moreover, on December~8, 2006, the wind depression profile was even observed directly in 
the He\,I~5876\,\AA{}. This is shown in Fig.\,13 together with the profile of the same line for September~15, 
2011, which has been shifted horizontally so that the cores coincidence. We find a deepening at 
Vr$\approx -50$\,km/s  on the blue slope of the first spectrum. Unfortunately, the December~8, 2006 spectrum 
does not include  the  long--wavelength  range containing the Si\,II~6347\,\AA{} and H$\alpha$ lines, but 
it does include  the H$\beta$ line, which has a weakly  expressed P~Cygni profile whose absorption component yields 
Vr\,$\approx -70$\,km/s. 

The differential line shifts and  profile shapes  line in the spectra for Cyg~OB2 No.\,12 are comparable 
to those for the hypergiants of similar spectral types mentioned above: HD\,80077~(B2.5), HD\,168625~(B5.5), and
HD\,183143~(B7)~[68]. The strong lines of all these stars display the asymmetry described above; the first
two exhibit positive shifts of the He\,I~5876\,\AA{} line core with respect to the weakest absorptions, 
though they are smaller than for No.\,12, and the amplitudes of the velocity time variations are also smaller. 
The profile asymmetry observed for HD\,80077 is especially striking. Several spectra of this southern star were 
kindly provided to us by G.A.~Galazutdinov. Even the weak absorption lines are clearly asymmetric in
them. Characteristic profiles of deeper lines in the spectrum, taken with the FEROS spectrograph~[69],
are shown in Fig.\,14. The resolution is close to our own, facilitating comparison of Fig.\,14 to Figs.\,11
and 13. The profile of the He\,I~5876\,\AA{} absorption in HD\,80077 is distorted by both a depression in its blue
wing and weak emission in its red wing. 

Combined data on radial velocities measured for various lines in the spectra of star No.\,12, HD\,80077, 
and HD\,168625 are collected in Table\,6. To facilitate comparison, we substracted Vsys from all the velocities; 
i.e. the table presents the differences  $\Delta$Vr\,=\,Vr$-$Vsys. Column~(2) contains velocity estimates 
derived from emission lines; only the first  estimate for star No.\,12 (2001.06.12) was made from the 
forementioned weak Fe\,II emission features near 7500\,\AA{}, while the later, less reliable estimates 
are based on the wide pedestals of the H$\alpha$ emission profiles. However, averaging results in a velocity 
that is consistent with Vsys\,=\,$-10.5\pm1$\,km/s for the association as a whole, confirming that this 
value can be used as the radial velocity of the star’s center of mass. Column~(3) presents the limiting
values to which the velocity differences converge in the transition from the strongest absorption 
features to weaker and weaker ones (r\,$\rightarrow 100$). The latter absorption features, mainly N\,II,
S\,II, Al\,III, and Si\,III lines, appear symmetric in the spectra of star No.\,12 and HD\,168625; this means 
that  their velocities were measured for the profile as a whole. In contrast, the velocities in columns~(4) 
and (5) are  based  {\bf on  the cores} of the asymmetric  lines Si\,II\,(2) and  He\,I~5876\,\AA{}  
(the estimates in brackets are for the  depressions  in the blue wings). 
The $\Delta$Vr estimates for the depressions and peaks in the H$\alpha$ profiles are given in columns (6) and (7).

Figure~15 presents H$\alpha$ profiles from 5 of our spectra. All of them, as well as the profiles presented
in~[65], are similar, and show strong bell--shaped emission lines with extended Thomson wings. The lower, symmetric 
part of the profile is more stable than the upper part. Variations are displayed mainly by the
central region, at Vsys\,$\pm 60$\,km/s. Intensity inversions are observed in this region in all the spectra, and only
one of these clearly also shows a depression in the blue wing that is separated from the profile center by
more than 100\,km/s. An attempt was made in~[65] to reproduce the observed H$\alpha$ profile of star No.\,12 using
a model with a homogeneous, spherically symmetric wind. The model with a limiting expansion velocity
of Vext\,=\,400 km/s, which was considered to provide the best fit, yields a standard P~Cygni profile that
reproduces well only the broad wings and the red slope of the main emission, while the blue part with its
typical absorption component is much lower than the blue slope of the observed profile. This is true even for
the spectrum of April~12, 2003 reproduced in Fig.\,11, in addition to the fact that the computed profile shows
no central inversion. 

The H$\alpha$ profiles in the spectra of HD\,80077 and HD\,168625 are  closer to the model ones; 
the absorption and emission  extrema are separated by 150--250\,km/s and are on opposite sides of the Vsys line. 
The extrema for star No.\,12 are closer  to each other, and both can occur on either side of the line. 
The wind is inhomogeneous: in addition to the continuously present,  relatively high-velocity portion, 
it contains a variable, sometimes considerable, amount of matter that is almost stationary with
respect to the star, or is even falling onto the star.

The presence of a slow component in the wind of star No.\,12 is the only reliably detected phenomenon
indicating that it is a member of the population of LBV stars, whose spectra (and even individual lines) can
display the coexistence of direct and inverse P~Cygni profiles. 

Generally speaking, our data do not exclude the possibility that star No.\,12 is a spectroscopic binary, 
but also do not confirm this hypothesis. Table~6 shows that the velocities change from epoch to epoch 
more strongly than the differential shifts of the absorption lines; at least partially, this could result 
from orbital motion. The broadened profile of the HeI~5876 line in Fig.\,13 could be obtained from the 
superposition of the spectra of the components of a two-spectrum, low-amplitude binary (SB2), but its 
$\gamma$--velocity would be too far from Vsys. In addition, other absorption lines in the spectrum 
of December~8, 2006 are not broadened. So far, no traces of a second component have been detected in 
the spectrum, and it is reasonable to consider only the SB1 option ---  a single  spectrum spectroscopic binary.

The anomalous reddening of star No.\,12, in excess of that for neighboring association stars, could originate 
either in the line of sight relatively far from the star or in its envelope. The data collected in
Table~5 tend to convince us in the second option. Figure~16 plots the equivalent widths of the D2\,NaI line and 
5797\,\AA{}  DIB from this table versus the color excess E\,(B--V) for association stars, supplemented
with analogous data for several foreground stars with E\,$< 1.2^m$. Star No.\,12 has repeatedly been included
in studies of interstellar lines, probably in the hope of finding them abnormally strong. However, Fig.~16
shows that their intensities do not exceed the level reached in the spectra of the most reddened stars, and
this is true not only for saturated Na\,I lines but also for DIBs remaining on the initial part of the curve of
growth.

\section{Conclusions}

The high-spectral resolution of the observations considered here made it possible to use weak lines,
enabling us to determine the main spatial and kinematic parameters of the association based on a small
number of its members. Our estimates of the distance and systemic velocity (d\,=\,1.7$\pm 0.1$\,kpc and 
Vsys\,=\,$-10.7\pm0.7$\,km/s) coincide, within the errors, with those derived from moderate  resolution spectra using
a much larger number of stars.

Moreover, we have spectroscopically confirmed age differences for stars representing different parts
of the association. The stars of the northern group (No.\,4,  6,  16,  21) are the oldest; from their 
positions in the HR--diagram, their ages are about 5 million years, and the available lines, including H$\alpha$, 
show no wind--related anomalies: they are symmetric absorption lines with minimal differential
shifts. The stars towards the southern boundary of the association’s central part (No.\,9, 22, 12, etc.)
are at least half as old, the absorption lines in their spectra are asymmetric, and the strongest of these
already display signatures of P~Cygni profiles. Information on the distribution and motion of cool
gas in the association’s center derived from our analysis of complex profiles of the Na\,I\,(1) doublet is 
likewise consistent with a scheme with cascade, possibly induced, star formation: the gas becomes denser from
north to south, and the gas outflow becomes slower and is accompanied by accretion in the vicinity of the
youngest stars.

Exploring this scheme in more detail for this scheme will require high-resolution spectroscopy for
additional stars, first and foremost, in the region of the southern clusters.
The possibility of comparing the spectra of stars located near each other in space is especially important
for refinding our understanding of the structure of the interstellar medium using the weak side components
of the Na\,I\,(1) lines; it is also possible to detect circumstellar features in this way (a transition from
wind lines to interstellar lines). 

Star No.\,12 (B4\,Ia) in the Cyg\,OB2 association is the only hypergiant, and the most evolved object 
among our program stars. Spectroscopic manifestations  of its unstable wind are quite clear and 
characteristic of early B--hypergiants -- not only the emission and absorption H$\alpha$ profiles, 
but also the characteristic asymmetry of the He\,I, Si\,II, and other lines,  as well as variations 
of their profiles with time and from line to line, up to a separation into photospheric and wind
components for the strongest of them. At the same time, the H$\alpha$ profile distinguishes star No.\,12 from
other known B hypergiants: it is formed in a slow wind that contains an accretion along with an excretion 
component.  This last feature can be considered evidence that this star is an LBV object.

However, the characteristic features of this class of variable stars  -- a specific combination of brightness,
color, and spectrum variations, an IR excess, and a dust circumstellar shell -- remain to be searched for,
as is also true of the star’s possible spectroscopic binarity. Such binarity could potentially reduce the
extreme luminosity of the star by dividing it between the system components, but in a very artificial way:
they would have to have very similar luminosities and also be virtually identical stars, since the spectrum
does not allow a difference in their temperatures.

Another argument of LBV status would be a circumstellar origin for the anomalous reddening of star No.\,12. 
However, thus far we have only indirect evidence for this: the interstellar Na\,I lines and DIBs
in the line of sight of star No.\,12 are not strong compared to the neighboring association members.

Having in mind the above findings, we will conclude by listing the main tasks for a further observational 
studies of Cyg\,OB2 and star No.\,12:

\begin{itemize}
 \item{Extending verified techniques to other association stars, beginning with those closest to already studied stars, and 
 to nearby H II regions.}
 \item{Obtaining higher resolution (R\,$\ge$100000) spectra for at least the brightest stars, to better
trace their Na\,I(1) profiles and stellar absorption lines. On the other hand, spectra with R\,$\approx$\,2000 
are needed to classify faint (V\,$\approx 14^m - 17^m$) stars in the immediate vicinity of star No.\,12.}
\item{The spectra we obtained are not sufficient to resolve the question of spectroscopic binarity
for star No.\,12, and thus monitoring is needed.}
\item{Speckle interferometry of star No.\,12 in the visible and near--IR is needed.}
\end{itemize}

\subsection*{Acknowledgements}

This study was supported by the Russian Foundation for Basic Research (project 11--02--00319\,a).  
The authors are grateful to G.A.~Galazutdinov for providing spectra of the star HD\,80077.

\bigskip

\section*{REFERENCES}
\begin{enumerate}

\item{}  L. M\"unch and W. W. Morgan, Astrophys. J. {\bf 118} 161 (1953).

\item{} H. L. Johnson and W. W. Morgan, Astrophys. J. {\bf 119}  344 (1954).

\item{} D. H. Schulte, Astrophys. J. {\bf 124} 530 (1956).

\item{}  D. H. Schulte, Astrophys. J. {\bf 128} 41 (1958).

\item{}  V. C. Reddish, L. C. Lawrence, and N. M. Pratt, Publ. R. Observ. Edinburgh  {\bf 5} 111 (1966).

\item{}  P. Massey and A. B. Thompson. AJ {\bf 101} 1408 (1991).

\item{}  J. Knodlseder, Astron. \& Astrophys. {\bf 360} 539 (2000).

\item{}  F. Comeron, A. Paqsuali, and G. Rodighiero, et al., Astron. \& Astrophys. {\bf 389} 874 (2002).

\item{}  M. M. Hanson, ApJ {\bf 597} 957 (2003).

\item{}  D. C. Kiminki, H. A. Kobulnicky, K. Kinemuchi, et al., ApJ {\bf 664} 1102 (2007).

\item{}  J. E. Drew, R. Greimel, M. J. Irwin, and S. E. Sale, MNRAS {\bf 386} 1761 (2008).

\item{}  J. S. Vink, J. E. Drew, D. Steeghs, et al., MNRAS  {\bf 387} 308 (2008).

\item{}  F. Comeron, A. Paqsuali, F. Figueras, and J. Torra, Astron. \& Astrophys. {\bf 486} 453 (2008).

\item{}  I. Negueruela, A. Marco, A. Herrero, and J. S. Clark, Astron. \& Astrophys. {\bf 487} 575 (2008).

\item{}  N. J. Wright, J. J. Drake, J. E. Drew, and J. S. Vink, ApJ {\bf 713} 871 (2010).

\item{}  C. Weidner, I. A. Bonnell, and H. Zinnecker, ApJ {\bf 724} 1503 (2010).

\item{}  J. S. Clark, I. Negueruela, B. Ritchie, et al., Messenger, No.\,142 31 (2010).

\item{}  E. Bica, Ch. Banatto, and C. M. Dutra, Astron. \& Astrophys. {\bf 405} 991 (2003).

\item{}  W. W. Morgan, H. L. Johnson, and N. G. Roman, PASP {\bf 66} 85 (1954).

\item{}  W. A. Hiltner, ApJ Suppl. Ser. {\bf 2} 389 (1956).

\item{}  S. Sharpless, PASP {\bf 69} 239 (1957).

\item{}  C. Leitherer, O. Stahl, B. Wolf, and C. Bertout, Astron. \& Astrophys. {\bf 140} 199 (1984).

\item{}  R. M. Humphreys and K. Davidson, PASP {\bf 106} 1025 (1994).

\item{}  A. M. van Genderen, Astron. \&  Astrophys. {\bf 366} 508 (2001).

\item{}  J. S. Clark, V. M. Larionov, and A. Arkharov, Astron. \& Astrophys. {\bf 435} 239 (2005).

\item{}  E. W. Gottlieb and W. Liller, ApJ {\bf 225} 488 (1978).

\item{}  P. R. Wozniak, W. T. Vestrand, C. W. Akerlof, et al., AJ {\bf 127} 2436 (2004).

\item{}  C. Sterken, T. Arentoft, H. W. Duerbeck, and E. Borgt, Astron. \& Astrophys. {\bf 349} 532 (1999).

\item{}  H. J. Wendker and W. J. Altenhoff, Astron. \& Astrophys. {\bf 92} L5 (1980).

\item{}  J. H. Bieging, D. C. Abbott, and E. B. Churchwell, ApJ {\bf 340} 518 (1989).

\item{}  G. E. Bromage, Nature {\bf 230} 172 (1971).

\item{}  S. P. Souza and B. L. Lutz, ApJ Lett. {\bf 235} L87 (1980).

\item{}  L. M. Hobbs, D. G. York, J. A. Thorburn, et al., ApJ {\bf 705}  32 (2009).

\item{}  V. G. Klochkova and E. L. Chentsov, Astron. Rep. {\bf 48} 1005 (2004).

\item{}  M. E. Contreras, G. Montes, and F. P. Wilkin, Rev. Mex. Astron. Astroﬁs. {\bf 40} 53 (2004).

\item{}  L. Luud, T. Tuvikene, and M. Ruuzalepp, Astroﬁzika {\bf 29} 97 (1988).

\item{}  A. Herrero, L. J. Corral, M. R. Villamariz, and E. L. Martin, Astron. \& Astrophys. {\bf 348} 542 (1999).

\item{}  A. Sota, J. Maiz Apellaniz, N. R. Walborn, et al., ApJ Suppl. Ser. {\bf  193} 24 (2011).

\item{}  N. R. Walborn, A. Sota, J. Maiz Apellaniz, et al., ApJ Lett. {\bf 711} L143 (2010).

\item{}  Y. Naze, M. De Becker, G. Rauw, and C. Barbieri, Astron. \&  Astrophys. {\bf 483} 543 (2008).

\item{}  A. Herrero, J. Puls, and M. R. Villamariz, Astron. \& Astrophys. {\bf 354} 193 (2000).

\item{}  V. E. Panchuk, V. G. Klochkova, M. V. Yushkin, and I. D. Najdenov, J. Optical Technology {\bf 76} 42 (2009).

\item{}  F. A. Musaev, G. A. Galazutdinov, A. V. Sergeev, et al., Kinem. Fiz. Nebesn. Tel  {\bf 15} 282 (1999).

\item{}  V. E. Panchuk, I. D. Najdenov, V. G. Klochkova, et al., Bull. Spec. Astrophys. Observ. {\bf 44} 127 (1997).

\item{}  M. V. Yushkin and V. G. Klochkova, Preprint Spets. Astroﬁz. Observ. No.206 (2005).

\item{}  G. A. Galazutdinov, Preprint Spets. Astroﬁz. observ. No.92 (1992).

\item{}  V. G. Klochkova, V. E. Panchuk, M. V. Yushkin, and D. S. Nasonov, Bull. Spec. Astrophys. Observ. {\bf 63} 410 (2008).

\item{}  S. Bagnulo, E. Jehin, C. Ledoux, et al., Messenger No.114, 10 (2003).

 \item{} G. A. Galazutdinov, F. A. Musaev, J. Krelowski, and G. A. H. Walker, PASP {\bf 112} 648 (2000).

 \item{} J. Maiz Apellaniz, Astron. \& Astrophys. {\bf 518} A1 (2010).

 \item{} M. De Becker, G. Rauw, and J. Manfroid, Astron. \& Astrophys. {\bf 424} L39 (2004).

 \item{} G. Rauw, J.-M. Vreux, and B. Bohannan, ApJ {\bf 517} 416 (1999).

 \item{} F. Royer, J. Zorec, and A. E. Gomez, Astron. \&  Astrophys. {\bf 463} 671 (2007).

 \item{} B. J. McCall, K. H. Hinkle, T. R. Geballe, et al., ApJ {\bf 567} 391 (2002).

 \item{} F. H. Chaffee and R. E. White, ApJ Suppl. Ser. {\bf 50} 169 (1982).

 \item{} R. Gredel and G. M\"unch, Astron. \& Astrophys. {\bf 285} 640 (1994).

 \item{} D. E. Welty and L. M. Hobbs, ApJ Suppl. Ser. {\bf 133} 345 (2001).

 \item{} L. M. Hobbs, ApJ {\bf 157} 135 (1969).

 \item{} S. E. Sale, J. E. Drew, Y. C. Unruh, et al., MNRAS {\bf 392} 497 (2009).

 \item{} K. Dobashi, Publ. Astron. Soc. Japan. {\bf 63} 1 (2011).

\item{}  N. Schneider, S. Bontemps, R. Simon, et al., Astron. \& Astrophys. {\bf 529} A1 (2011).

\item{}  N. Schneider, S. Bontemps, R. Simon, et al., Astron. \& Astrophys. {\bf 458} 855 (2006).

\item{}  T. G. Sitnik, T. A. Lozinskaya, and V. V. Pravdikova, Astron. Lett. {\bf 33} 814 (2007).

\item{}  Y. M. Butt, N. Schneider, T. M. Dame, and Ch. Brunt, ApJ Lett. {\bf 676} L123 (2008).

\item{}  J. S. Clark, F. Najarro, I. Negueruela, et al., ApJ {\bf 541} 145 (2012).

\item{}  V. G. Klochkova, E. L. Chentsov, N. S. Tavolzhanskaya, and G. A. Proskurova, Preprint Spec. Astrophys. 
         Observ. No.183 (2003).
         
\item{}  G. Knoechel and A. F. J. Moffat, Astron. \& Astrophys. {\bf  110} 263 (1982).

\item{}  E. L. Chentsov, Astron. Lett. {\bf 30} 325 (2004).

\item{}   A. Kaufer, O. Stahl, S. Tubbesing, et al., Messenger {\bf 95} 8 (1999).
\end{enumerate}

\newpage

\begin{table}
\bigskip
\caption{\small  Log of spectra used}
\begin{tabular}{l  r  c  c  c  l}
\hline
Star&  Date  &    Spectrograph  & R  & $\Delta\lambda$,\,\AA{} & Reference \\ [-3pt]
\hline 
\multicolumn{6}{l}{\small\underline{Stars in Cyg~OB2:}} \\[-1pt]
No.\,2 &   25.07.07&    NES    &60000  &4555--6015 &[6] \\[-1pt]
No.\,4 &    1.09.04&    NES    &60000  &5275--6770 &[37, 38] \\ [-1pt]         	 	     	  
No.\,5 &   31.08.97&   MAESTRO &40000  &3900--8630 &[38] \\ [-1pt]
       &   14.11.97&   MAESTRO &40000  &3900--9240	&  \\ [-1pt]
       &   14.09.11&    NES    &60000  &4100--6900	&  \\ [-1pt]
No.\,21&   18.11.10&    NES    &60000  &5215--6690 &[10, 37]\\[-1pt]
No.\,16&   10.06.11&    NES    &60000  &4850--6240 &[6] \\[-1pt]
No.\,12&   12.06.01&    PFES   &15000  &4900--7780 &[10, 34] \\ [-1pt] 
       &   12.04.03&    NES    &60000  &5270--6770	&\\ [-1pt]
       &    8.12.06&    NES    &60000  &4470--5940	&\\ [-1pt]
       &   26.09.10&    NES    &60000  &5210--6690	&\\ [-1pt]
       &   19.11.10&    NES    &60000  &5210--6690	&\\ [-1pt]
       &   15.09.11&    NES    &60000  &4900--6900	&\\ [-1pt]
No.\,6 &    3.06.10&    NES    &60000  &5300--6690 &[6]  \\ [-1pt]
       &    4.06.10&    NES    &60000  &4465--5930	&\\ [-1pt]
No.\,22&   10.06.11&    NES    &60000  &4850--6240 &[37, 38] \\ [-1pt]
No.\,9 &    3.08.09&    NES    &60000  &4400--5860 &[37, 39, 49] \\ [-1pt]
       &   31.07.10&    NES    &60000  &4465--5930	&\\ [-1pt]
No.\,7 &    9.06.11&    NES    &60000  &4850--6240 &[38, 41]  \\ [-1pt]
No.\,8А&   30.09.99&   MAESTRO &40000  &3800--9240 &[6, 39] \\ [-1pt]
       &   30.08.04&    NES    &60000  &5300--6710	&\\ [-1pt]
No.\,18&   24.07.07&    NES    &60000  &4555--6015 &    \\ [-1pt]
No.\,10&   28.08.04&    NES    &60000  &5300--6710 & [37] \\ [-1pt]
\multicolumn{6}{l}{\small\underline{Foreground star BD\,+41\degr3814}}\\ [-1pt]
       &   12.01.11&    NES    &60000  &5200--6680	& \\  
\hline
\multicolumn{6}{l}{\it\footnotesize \underline{Note:} The first column presents  the star numbers from [4]}\\
\end{tabular}   
\end{table}

  \begin{table}
  \bigskip
  \caption{\small  Data on program stars}
  \begin{tabular}{ l  c   r  l r l r  r  l }
  \hline
  & $\alpha, \delta_{2000}$ & Sp & Mv &V  &B--V & Av  & d \\ 
  &                          &    &mag &mag&mag & mag &  кпк \\
  \hline
  \multicolumn{6}{l}{\small\underline{Stars in Cyg~OB2}}\\ 
  No.\,7 &  20$^h$33$^m$14$^s$  & O3\,If &$-6.4$ & 10.55 & 1.45 &  5.3 &  2.0 \\ 
         &  41\degr20\farcm22\farcs  &&&&&& \\ [-1pt]
  No.\,22&  20$^h$33$^m$08$^s$  & O3\,I  &$-6.5$:& 11.55 & 2.04 &  7.0 &  1.8:\\ 
         &  41\degr13\farcm18\farcs  &(+ O6 V)&&&&& \\ [-1pt]
  No.\,9 &  20$^h$33$^m$11$^s$  &O4.5\,If&$-6.7$ & 10.96 & 1.81 &  6.4 &  1.7 \\ 
         &  41\degr15\farcm08\farcs  &&&&&& \\ [-1pt]
  No.\,8A&  20$^h$33$^m$15$^s$  &O5\,III &$-6.8$ & 9.06  & 1.30 &  4.9 &  1.5 \\ 
         &  41\degr18\farcm50\farcs  &&&&&& \\ [-1pt]
  No.\,4 &  20$^h$32$^m$14$^s$  &O7\,IIIf&$-5.6$ & 10.23 & 1.19 &  4.5 &  1.8 \\ 
         &  41\degr27\farcm12\farcs  &&&&&& \\ [-1pt]
  No.\,5 &  20$^h$32$^m$22$^s$  &O7\,If  &$-6.9$:&  9.21 & 1.43 &  5.2 &  1.6: \\
         &  41\degr18\farcm19\farcs  &&&&&& \\ [-1pt]
  No.\,16& 20$^h$32$^m$39$^s$  &O7.5\,V &$-5.0$ & 10.84 & 1.19 &  4.5 &  1.8 \\ 
         &  41\degr25\farcm14\farcs  &&&&&& \\ [-1pt]
  No.\,6 &  20$^h$32$^m$45$^s$  &O8\,V   &$-4.8$ & 10.68 & 1.25 &  4.7 &  1.5 \\
         &  41\degr25\farcm37\farcs  &&&&&& \\ [-1pt]
  No.\,21& 20$^h$32$^m$28$^s$  &B0\,V   &$-3.8$ & 11.42 & 1.00 &  3.9 &  1.8 \\ 
         &  41\degr28\farcm52\farcs  &&&&&& \\ [-1pt]
  No.\,12& 20$^h$32$^m$41$^s$  &B5\,Ia-0&$-9.5$:& 11.46 & 3.30 & 10.0 &  1.6: \\ 
         &  41\degr14\farcm29\farcs  &&&&&& \\ [-1pt]
  \multicolumn{6}{l}{\small\underline{Маловероятные члены Cyg\,OB2}}\\
  No.\,10& 20$^h$33$^m$46$^s$  &O9.5\,Iab&$-6.4$&  9.88 & 1.59 &  5.6 &  1.4 \\ 
         &  41\degr33\farcm01\farcs  &&&&&& \\ [-1pt]
  No.\,2 &  20$^h$31$^m$22$^s$  &B0.5\,II &$-5.1$& 10.64 & 1.20 &  4.3 &  2.0 \\
         &  41\degr31\farcm28\farcs  &&&&&& \\ [-1pt]
  No.\,18& 20$^h$33$^m$31$^s$  &B1\,Ib   &$-5.8$& 11.01 & 1.77 &  5.9 &  1.5 \\ 
         &  41\degr15\farcm23\farcs  &&&&&& \\ [-1pt]
  \multicolumn{6}{l}{\small\underline{BD\,+41\degr3814}}\\ 
         & 20$^h$34$^m$57$^s$  &B8\,V    & +0.1  &  7.59 &$-0.10$ & 0.0 &  0.3 \\ 
         &  41\degr43\farcm04\farcs  &&&&&& \\ 
  \hline 
  \multicolumn{8}{l}{{\it \footnotesize \underline{Notes:}  The number in the first column is from [4]. Star No.\,22 }}\\
  \multicolumn{8}{l}{{\it \footnotesize is a close visual double with a component separation of 1.5 arc.sec and }}\\  
  \multicolumn{8}{l}{{\it \footnotesize a component magnitude difference of 0.6$^m$ [50]. The fourth column}} \\
  \multicolumn{8}{l}{{\it\footnotesize  presents the spectral types  of the primary (and, in one single case, }}\\
  \multicolumn{8}{l}{{\it\footnotesize  secondary) components  according to [38]. The components were not resolved }} \\ 
  \multicolumn{8}{l}{{\it \footnotesize in our spectra, but the contribution from the  secondary  is appreciable:}}\\
  \multicolumn{8}{l}{{\it\footnotesize  we clearly see N\,IV 5200-–5205\,\AA{}  absorbtion and 6215, 6220\,\AA{}}} \\  
  \multicolumn{8}{l}{{\it\footnotesize  emission lines, and the  absorbtion and  emission lines are weaker  than in}} \\ 
  \multicolumn{8}{l}{{\it\footnotesize  the  spectrum of star No.\,7; the C IV 5801, 5812\,\AA{}  lines appear in }} \\ 
  \multicolumn{8}{l}{{\it\footnotesize  emission in the  spectrum of star No.\,7 and are absent from the spectrum  }} \\
  \multicolumn{8}{l}{{\it\footnotesize  of star No.\,22; on the contrary, the C\,III~5696\,\AA{}  emission is }}\\
  \multicolumn{8}{l}{{\it\footnotesize  not visible for the former star but is visible  for the latter (Fig.\,1.}}\\
  \end{tabular}   
  \end{table}

\begin{table}
\bigskip
\caption{\small  Mean heliocentric radial velocities Vr for groups of lines and individual lines in the spectra of Cyg~OB2 stars
         other than star No.\,12}
\begin{tabular}{ l  r r l r  r  l  r  r  c c  l  l l}
\hline
&&\multicolumn{9}{c}{\small Vr, km/s}\\ 
\cline{3-11}
  Star  & $\Delta$Vr&Emis &$r\approx 100$& HeI &H$\beta$ &H$\alpha$ & NaI\,(1)& DIB & systemic &other studies & Ref  \\  
\hline
    1    &    2   & 3 &  4   &  5  & 6  &  7  & 8   &  9    &  10 &  11 & 12  \\
\hline
No.\,7 &    90  &--13 & --10 &--12P &--13P&      &--7.4&  --11 &--11 & --28& [10] \\ 
No.\,22&   120: &--12 & --12 &--20: &--16P&      &--9.0&  --10 &--12 &     &       \\  
No.\,9 &   130  & --7 &  --8 &--10P &--30P&--200P&--7.9&  --10 & --8 & --16& [10, 40]\\ 
No.\,8A&        &--56 & --55:&--65: &--60 & --67 &--9.0&  --11 &     & --68& [51] \\
       &        &     &  +40:& +40: &     &      &     &       &     &  +40&       \\     
No.\,4 &   130  &--12 &  --7 &--13  &     & --18 &--8.1&  --10 & --8 &  --6& [10] \\
No.\,5 &        &     &  +42 &      &     &      &--8.6&  --11 &     &  +40& [52]\\
       &        &     &--250:&      &     &      &     &       &     &--265& (31.08.97)\\ 
No.\,5 &        &     & --95 &      &     &      &--8.4&  --11 &     &--105& (14.11.97)\\ 
       &        &     & +190:&      &     &      &     &       &     & +200& \\	          
No.\,5 &        &     &      &      &     &      &--8.2&  --10 &     &     &(14.09.11) \\ 
No.\,16&   220  &     & --11:&--13: &--19 &      &--9.4&  --10 &--11:& --15&[10] \\
No.\,6 &   230: &     & --11 &--11  &--19 & --17 &--5.7&  --10 &--11 & --12&[10] \\
No.\,10&    90  &--10 &  --9 &--18  &     &--165P&--9.0&  --10 & --9 &--5: &[10] \\
No.\,21&    30  &     & --14 &--16  &     & --16 &--8.0&  --11 &--14 & --14&[10] \\
No.\,2 &    55  &     &  --6 & --6  & --4 &      &--8.6&   --8 & --6 &  --5&[10] \\
No.\,18&    95  &     &  --5 & --5  & --2 &      &--8.2&   --9 & --5 &  --6&[10] \\
\multicolumn{11}{l}{\small\underline{BD\,+41\degr3814}} \\ 
       &   120  &     & --16 &--11  &     & --12 &--15.2&   --16:& --15 &     &  \\	       
\hline	        								         
\multicolumn{12}{l}{{\it\footnotesize \underline{Note:} The first column is the star number from [4] and the second column the line 
           width $\Delta$Vr in km/s.}} \\
\end{tabular}    
\end{table}

\begin{table}
\bigskip
\caption{\small  Variation of heliocentric radial velocities Vr for components of the interstellar D2~Na\,I\,(1) in the spectra of 
           Cyg~OB2 stars with the line depth R}
\begin{tabular}{l  r  r r r  r  r  r l}
\hline
 Star & l, degree &   V$_{\rm blue}$  & \multicolumn{5}{c}{\small Vr} & Remarks   \\ 
    &   b, degree &   V$_{\rm red}$  & \multicolumn{5}{c}{\small R}   &   \\ 
\hline
No.\,12&  80.10& --19.3 &      &    & --33  &  +21 &   +34  &  K\,I [54, 55] \\ 
       &   0.83&   +1.5 &      &    &   64  &   19 &     9  &  C$_2$ [54, 56]  \\ 
       &       &        &      &    &       &      &        &  CO [54]      \\ 
No.\,5 &  80.12& --17.8 &      &    & --34  &  +30:&   +41: &  K\,I [54, 55] \\ 
       &   0.91&   +1.0 &      &    &   67  &   11:&    18: &  C$_2$ [54, 56] \\ 
       &       &        &      &    &       &      &        &  CH [54] \\ 
No.\,22&  80.14& --19.0 &      &    & --32  &  +19:&        & \\ 
       &   0.75&   +2.0 &      &    &   60: &   10 &	   & \\ 
No.\,9 &  80.17& --18.5 &      &    & --31  &  +12 &   +27  &  K\,I [55] \\ 
       &   0.76&   +2.0 &      &    &   73  &   66 &    19  & \\ 
No.\,2 &  80.18& --17.0 &  --50&    & --28: &  +17 &        & \\ 
       &   1.20&  --1.8 &    15&    &   52: &    5 &	   & \\ 
No.\,18&  80.21& --18.3 & --51:&    & --38: &      &        & \\ 
       &   0.71&   +2.2 &   15 &    &   34: &      &	   & \\ 
No.\,8A&  80.22& --19.1 &  --49&    & --29: &      &        &K\,I [55] \\ 
       &   0.79&   +2.0 &   17 &    &   52: &      &	   & \\ 
No.\,7 &  80.24& --17.2 & --54:&    & --29: &  +14:&        & \\
       &   0.80&   +2.1 &    9 &    &   26: &   35 &	   & \\ 
No.\,16&  80.24& --18.7 &  --47&    & --29: &  +19 &        & \\
       &   0.94&   +0.6 &   18:&    &   44: &   19 &        & \\
No.\,6 &  80.26& --17.3 &  --50&    & --29: &  +20 &        & \\
       &   0.93&   +5.5 &    25&    &   32: &   30 &	   & \\ 
No.\,4 &  80.22& --18.1 &  --49&    & --33: &  +18 &        & \\
       &   1.02&   +1.2 &    11&    &   16: &   27 &	   & \\ 
No.\,21&  80.27& --18:  &  --53&--46& --28: &  +19 &        & \\
       &   1.01&   +2:  &    14&  19&   75: &   20 &	   & \\ 
No.\,10&  80.47& --19.8 & --50:&    & --34: &      &        &K\,I [55] \\ 
       &   0.85&   +1.5 &    3:&    &   8   &      &	   & \\ 
\hline
\multicolumn{9}{l}{\small\underline{\,BD+41\degr3814}} \\ 
      &  80.74 &$-21.2$ &&&&&& \\  
      &   0.77 &$-10.7$ &&&&&& \\
\hline	        								       
\multicolumn{9}{l}{{\footnotesize {\it Note.}  The first column gives the star number from [4].}} \\
\end{tabular}   
\end{table}

\begin{table}
\bigskip
\caption{\small Equivalent widths W and depths R for interstellar lines and bands. The star numbers in the first column are from [4].} 
\begin{tabular}{l  c| l l l l l l l r}                                             
\hline
    & W(D2)$/$&&\multicolumn{7}{c}{\small W$/$R for DIB}\\ 
    &  W(D1)  &$\lambda$= & 5494 &  5780  & 5797 &  5849 &  6196 &  6379 &  6613 \\ [0pt]    
\hline	       	 
No.\,12&   0.84  & & 0.06 &  0.88 &  0.32 &  0.13 &  0.11 &  0.16 &  0.38  \\
       &   0.75  & &  10  &    37 &   31 &    15  &    20 &    22 &    32  \\
No.\,5 &   0.80  & & 0.04 &  0.80 &       &  0.09:&  0.09 &  0.10 &  0.30  \\
       &   0.67  & &  06  &    33 &       &    10:&    17 &    15 &    27  \\
No.\,22&   0.84  & & 0.07 &  0.99 &  0.29 &  0.08 &  0.11 &       &        \\
       &   0.70  & &  08  &    44 &    29:&    10:&    21 &       &        \\
No.\,9 &   0.94  & & 0.07 &  1.02 &  0.31:&  0.14 &       &       &        \\
       &   0.83  & &  09: &    40 &    30:&    14 &       &       &        \\
No.\,2 &   0.67  & & 0.04:&  0.79 &  0.24 &  0.09:&       &       &       \\ 
       &   0.55  & &   06 &    32 &    21 &    08 &       &       &       \\ 
No.\,18&   0.82  & & 0.06:&  1.03 &  0.33:&  0.12 &       &       &       \\ 
       &   0.69  & &   08:&    42 &    29 &    12 &       &       &       \\ 
No.\,8A&   0.79  & & 0.06 &  0.85 &  0.20 &  0.09 &  0.10 &  0.12 &  0.34 \\ 
       &   0.67  & &   08 &    34 &    21 &    09 &    18 &    16 &    29 \\ 
No.\,7 &   0.74  & & 0.06:&  0.80 &  0.24 &  0.10 &  0.09 &       &       \\ 
       &   0.64  & &   07 &    33 &    23 &    10 &    17 &       &       \\ 
No.\,16&   0.80  & & 0.03 &  0.78 &       &  0.07 &  0.09 &       &       \\ 
       &   0.63  & &   04 &    33 &    18:&    07:&    18 &       &       \\ 
No.\,6 &  0.89   & &0.04: & 0.79  &  0.21 & 0.06: & 0.09  &  0.07 & 0.30  \\ 
       &   0.75: & &   05 &    33 &   18 &     07 &    16 &    21 &   24  \\ 
No.\,4 &   0.68: & & 0.05:&  0.68:&  0.16:&  0.06:&  0.08 &  0.09:&  0.30 \\ 
       &   0.61: & &   07:&    30 &    19 &    08:&    15:&    13:&    25 \\ 
No.\,21&   0.89  & & 0.05 &  0.78 &  0.18:&  0.07:&  0.08 &  0.08:&  0.28 \\ 
       &   0.76  & &   06 &    31 &    17 &    05 &    15 &    10 &    23 \\ 
No.\,10&   0.64  & & 0.06:&  1.08 &  0.21 &  0.07:&  0.11 &  0.10 &  0.38 \\ 
       &   0.57  & &   07 &   45 &     19 &    07 &   20 &     14 &    31 \\ 
\hline
\multicolumn{9}{l}{\small\underline{BD\,+41\degr3814}} \\  
    &   0.13  & &      & 0.04: &       &       & 0.01: &       &    \\
    &   0.07  &&&&&&&& \\
\hline	        								       
\end{tabular}   
\label{DIB} 
\end{table}

\begin{table}
\bigskip
\caption{Differences between the observed heliocentric radial velocities and the systemic velocity, 
         $\Delta$Vr\,=\,Vr$-$Vsys,  for some lines in the spectra of Cyg~OB2 No.\,12, HD\,80077, and HD\,168625}
\begin{tabular}{ r| c  c c  c c c }
\hline
  Дата &\multicolumn{6}{c}{$\Delta$Vr\,=\,(Vr$-$Vsys), km/s} \\
\cline{2-7}  
       & Emis  & r\,$\approx 100$ & SiII\,(2)& HeI & H$\alpha$--abs & H$\alpha$--emis   \\
\hline
\multicolumn{7}{l}{\small \underline{CygOB2~No.\,12}}\\
12.06.01 &0     &$-5 $ &$-4$  &$8 $      &$ 10$       & $-40$:, 55  \\
12.04.03 &$-4$: &$-14$ &$-18$ &$0$       &$-113, -17$ & $-42$, $-22$\\ 
 8.12.06 &      &$-14$ &      &5($-40:$) &            &             \\ 
26.09.10 &$1  $ &$-1 $ &$-6$  &$ 11$     &$ 29$       & $ 33$       \\
19.11.10 &$4$:  &$-10$:&$-9$  &13($-25$):  &$-110$:, $10$:& $-28$       \\
15.09.11 &$-9$: &$-3$  &$-5$  &$-3$      &$-42$       & $-65$:, $25$   \\
\hline
\multicolumn{7}{l}{\small \underline{HD\,80077}}\\
         &      &$-2$  &2($-50:$) &$-10(-92)$& $-122$ & 34 \\ 
\hline
\multicolumn{7}{l}{\small \underline{HD\,168625}}\\
        & 2:    & 2 & $-5$ & $-2$ & $-170$ & 50 \\  
\hline
\end{tabular}   
\end{table}

\clearpage

\newpage

\begin{figure}
\hbox{
\includegraphics[angle=0,width=0.5\textwidth,bb=35 30 550 780,clip]{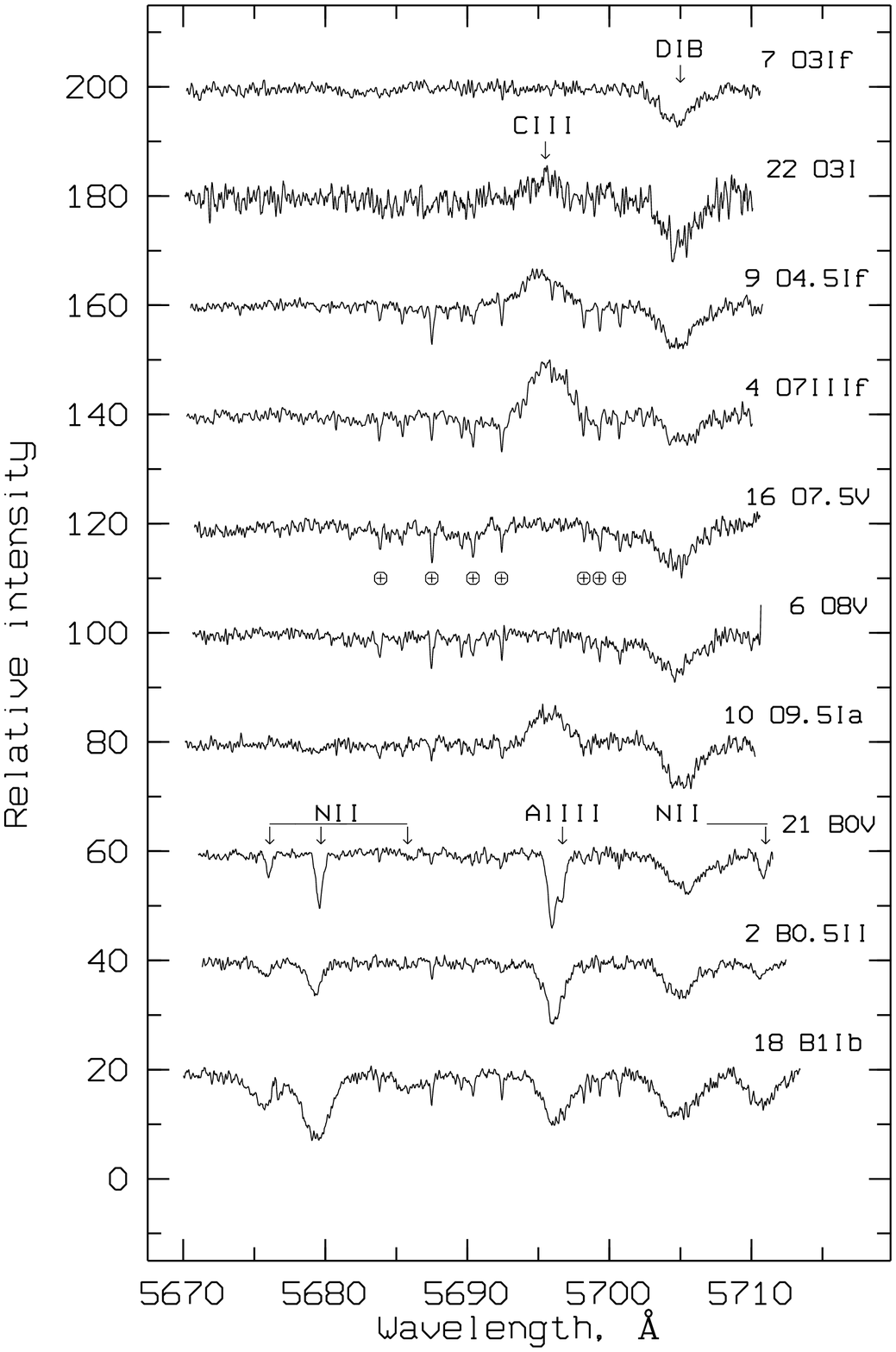}   
\includegraphics[angle=0,width=0.4\textwidth,height=0.5\textheight,bb=135 50 550 780,clip]{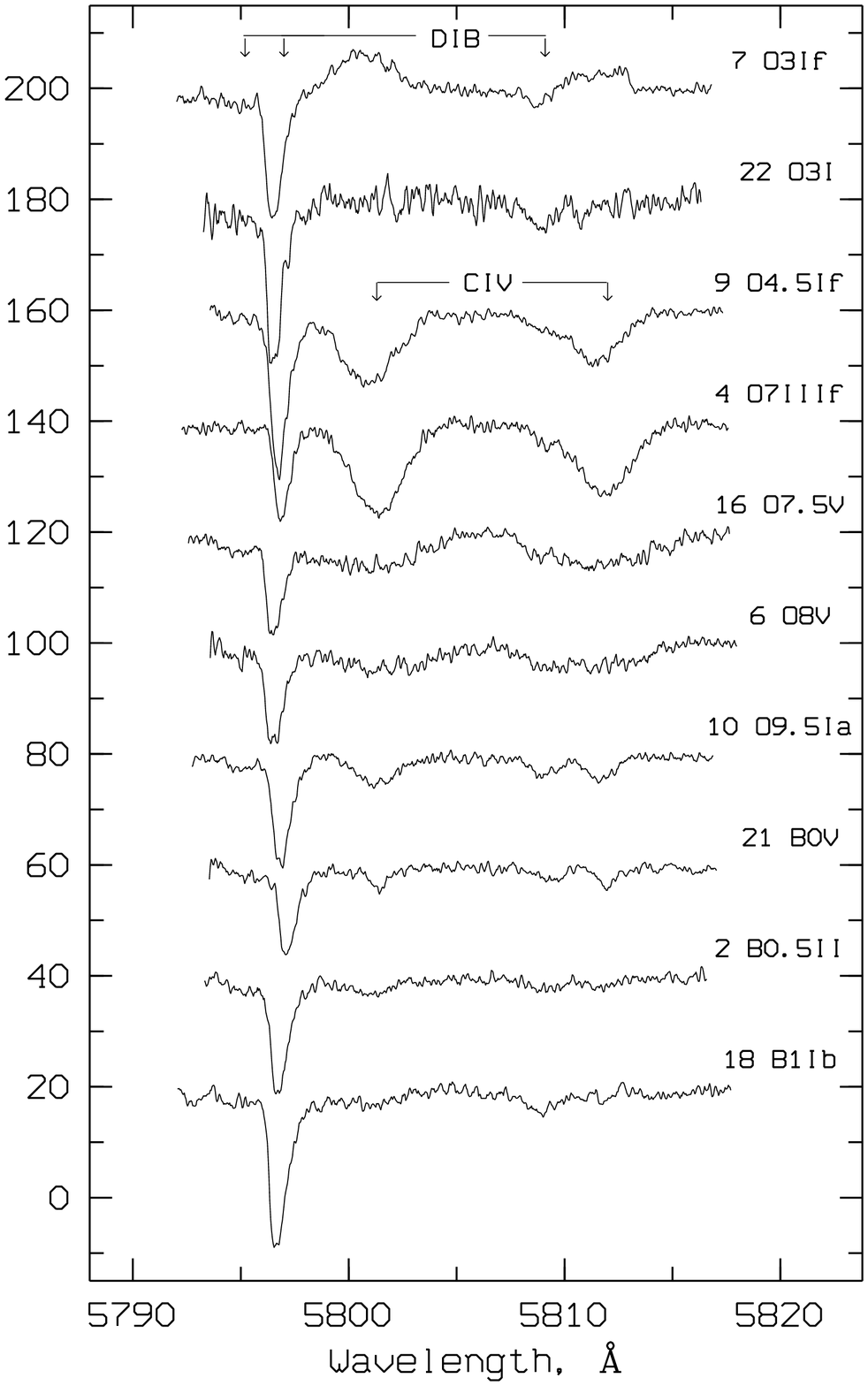} 
}
\caption{Fragments of spectra of Cyg~OB2 stars.  The stellar spectra are  shifted relative to each other 
         by 20 continuum units. The star’s spectral type is indicated near its number. 
         Left: the wavelength range containing the C\,III(2)~5695.92\,\AA{}, N\,II(3) 5676.02, 5679.56, 5686.21 
         and 5710.77\,\AA{}  and Al\,III 5696.60\,\AA{} stellar lines and the 5705.20\,\AA{} DIB.   
         The circled crosses mark, the strongest H$_2$O telluric lines in this fragment.  
         Right: the wavelength range containing the C\,IV(1) 5801.33, 5811.98\,\AA{} 
         stellar doublet and the 5795.16, 5796.96, and 5809.24\,\AA{} DIBs.}
\end{figure}

\begin{figure}
\includegraphics[angle=0,width=0.7\textwidth,bb=35 30 550 780,clip]{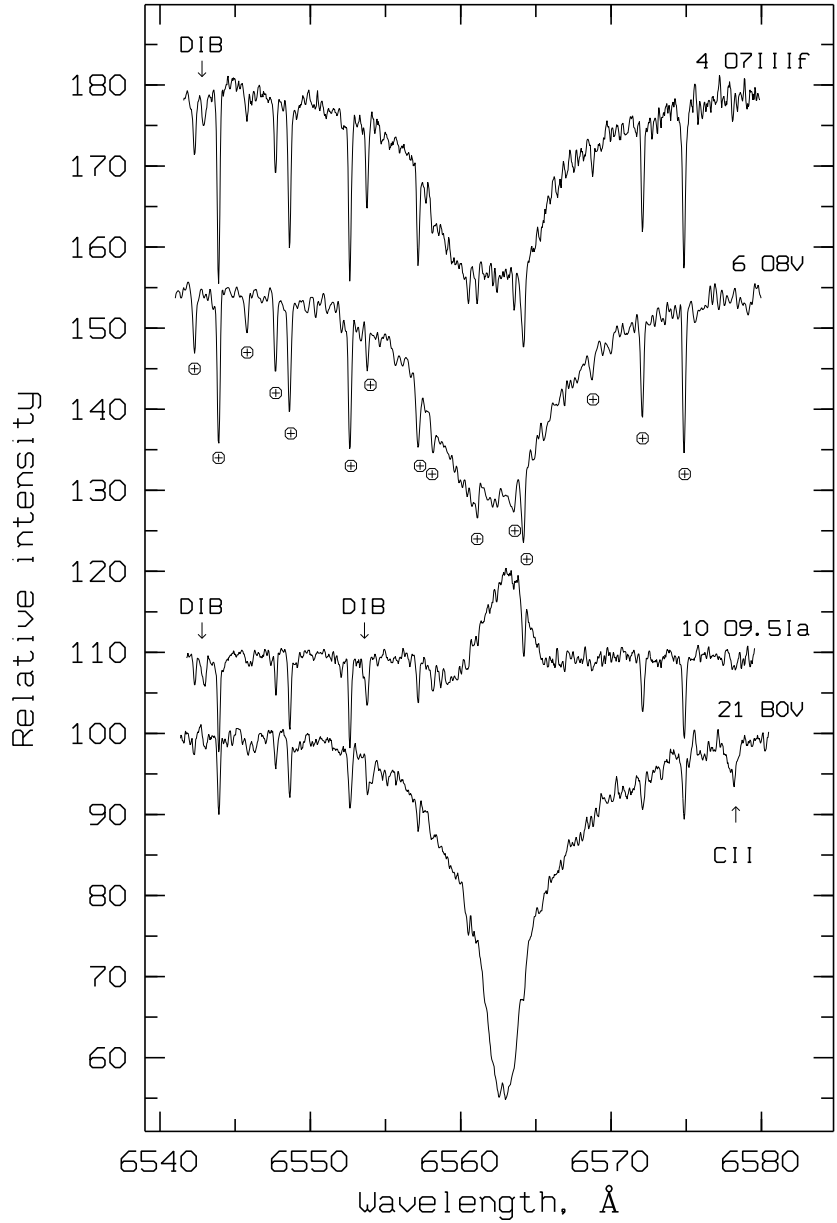} 
\caption{Fragments of the available spectra of 4 Cyg~OB2 stars in the region of H$\alpha$. 
         A star’s spectral type is indicated near its number. The arrows mark the narrow  6543.20 and 6553.82\,\AA{} DIBs ; 
         the circled crosses mark  the telluric H$_2$O lines at 6542.3,  6543.9, 6545.8, 6547.7, 6548.7, 6552.7, 
         6554.0, 6557.3, 6558.1, 6561.1, 6563.6, 6564.4, 6568.8, 6572.1, and 6574.9\,\AA{}.} 
\end{figure}

\begin{figure}
\hbox{
\includegraphics[angle=0,width=0.35\textwidth,height=0.38\textheight,bb=35 30 550 780,clip]{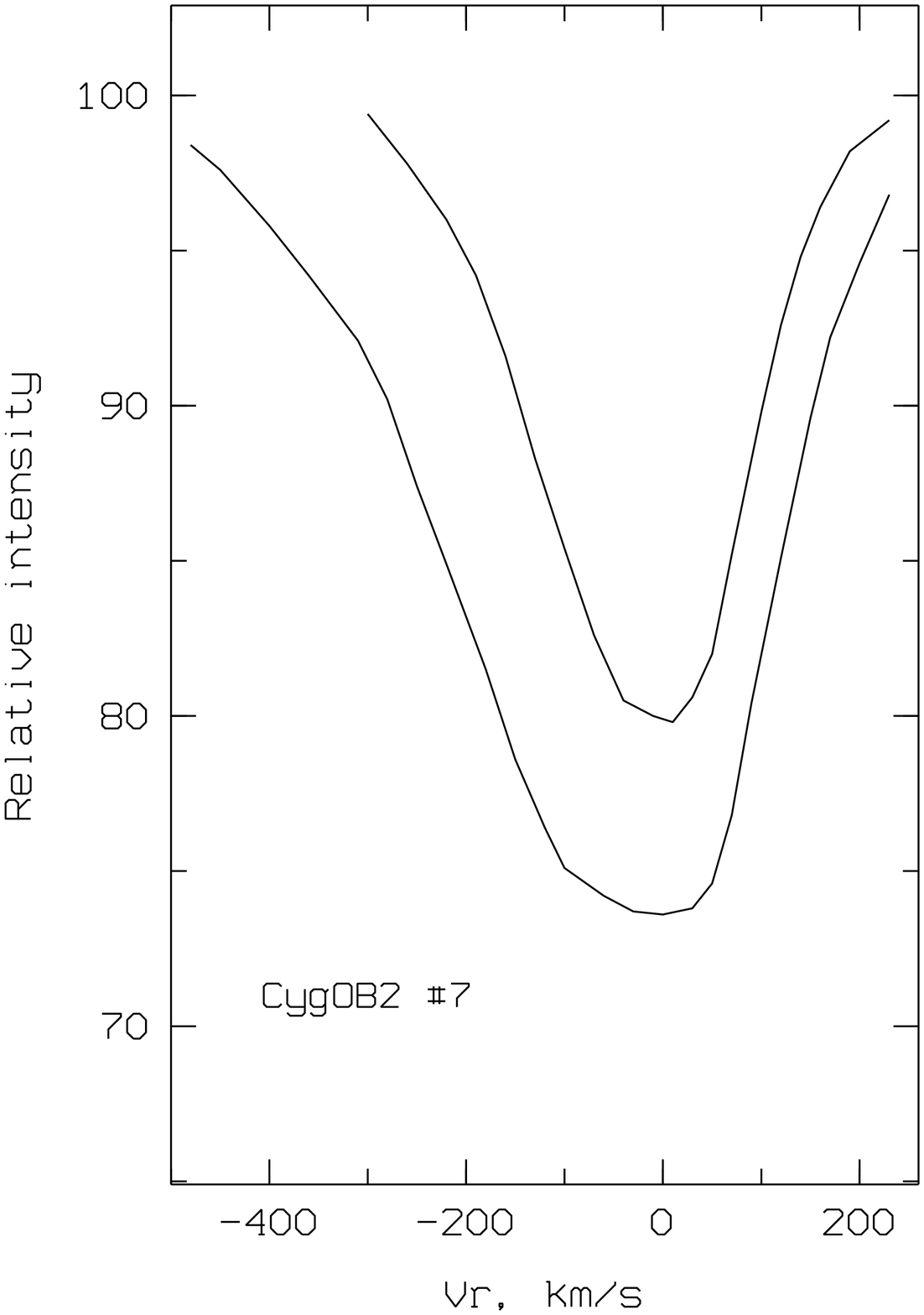}  
\includegraphics[angle=0,width=0.30\textwidth,height=0.38\textheight,bb=75 30 550 780,clip]{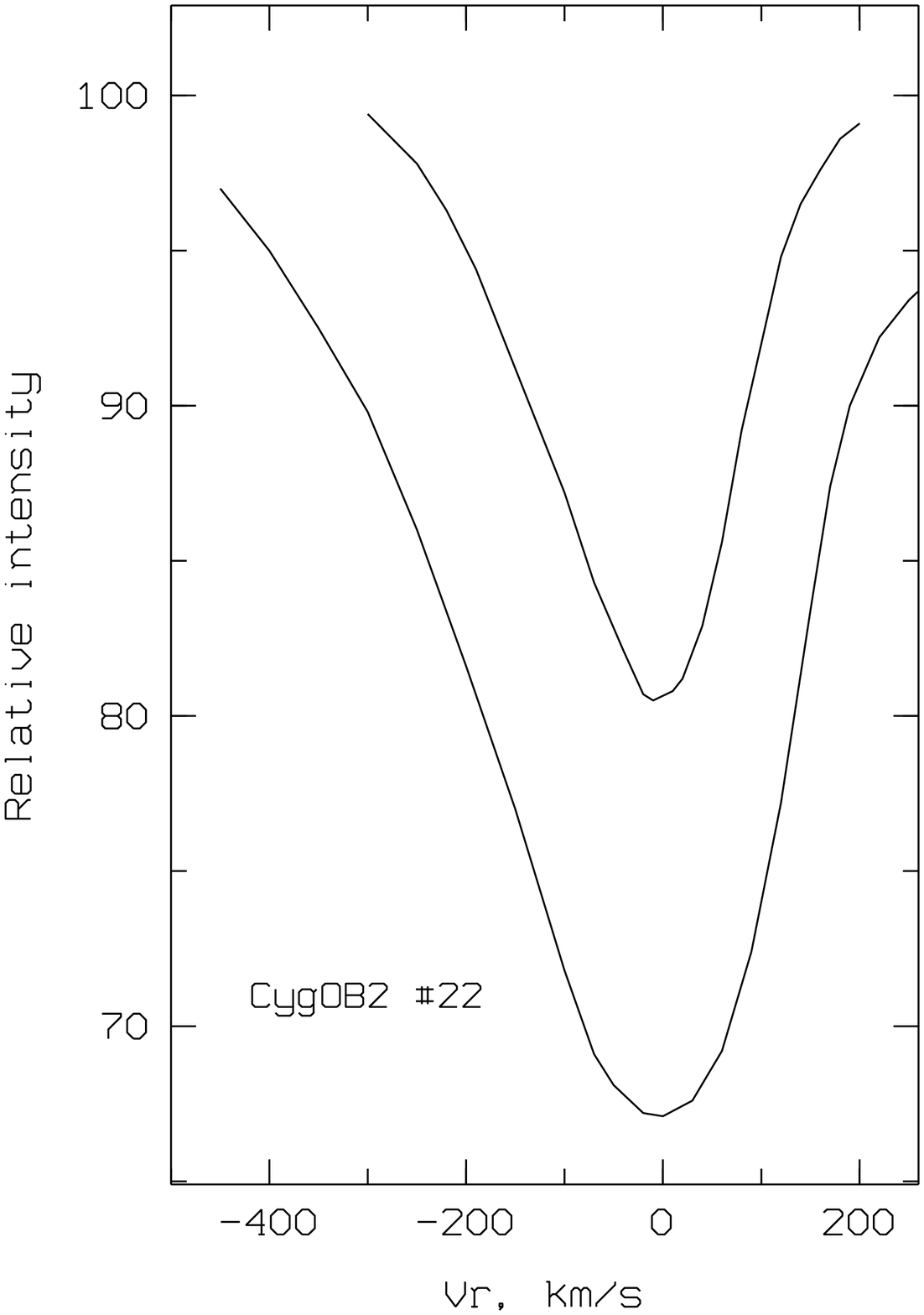}  
\includegraphics[angle=0,width=0.30\textwidth,height=0.38\textheight,bb=75 30 550 780,clip]{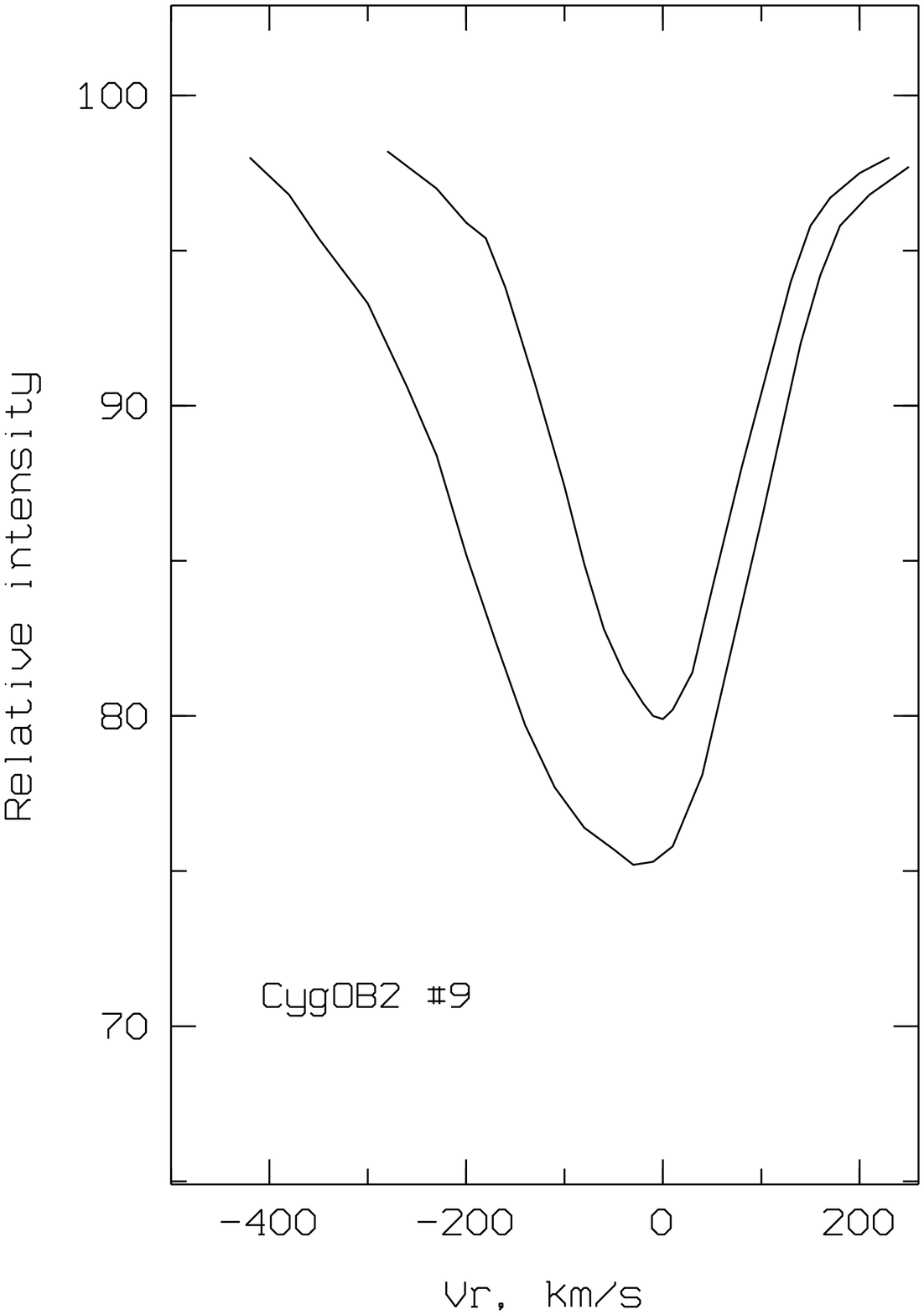}   
}
\caption{Smoothed asymmetric profiles of the H$\beta$ (lower curves) and He\,II~5411.52\,\AA{}  (upper curves) lines in the spectra of stars 
            Nos.\,7~(O3\,If), 22~(O3\,I + O6\,V), and 9~(O4.5\,If).}
\end{figure}

\begin{figure}
\hbox{
\hspace{-1.5cm}
\includegraphics[angle=-90,width=0.6\textwidth,bb=30 60 550 780,clip]{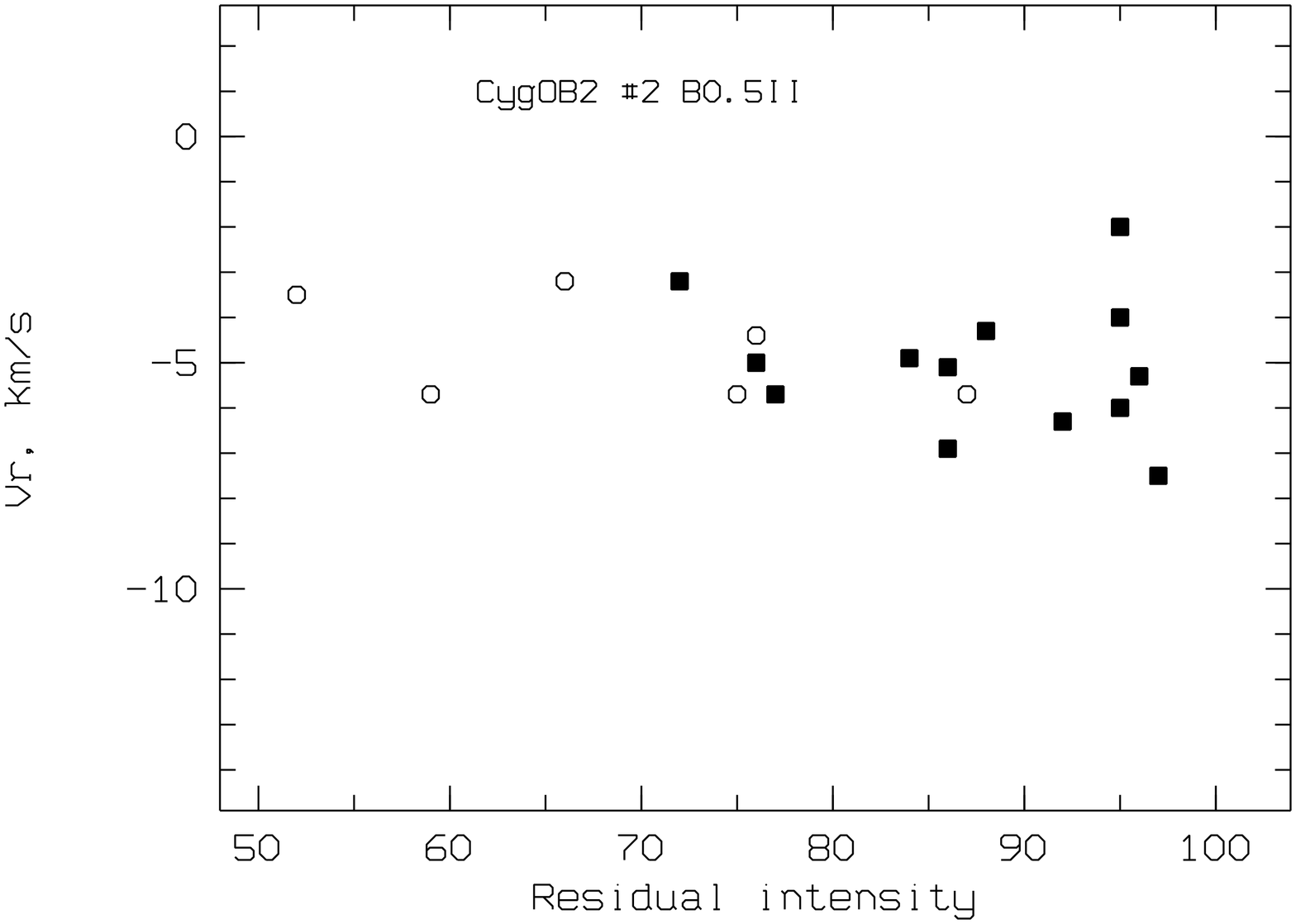} 
\includegraphics[angle=-90,width=0.5\textwidth,bb=30 180 550 780,clip]{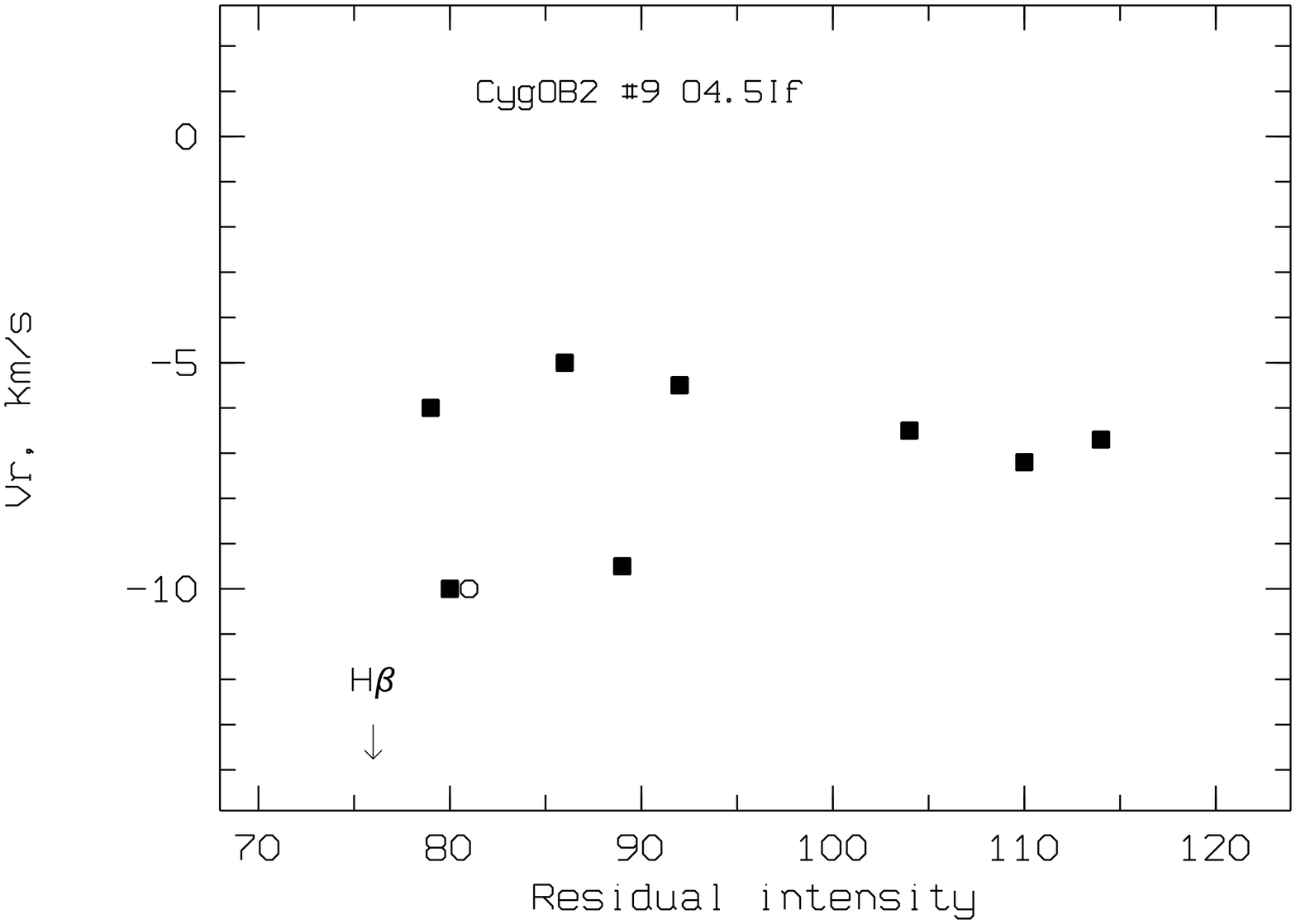} 
}
\caption{Dependences between the heliocentric radial velocity and the line residual intensity Vr\,(r) for Cyg~OB2 stars No.\,9
         and No.\,2. The velocities for star No.\,9 were measured from absorption and emission features, and those for star 
         No.\,2 from absorption features only. Each symbol corresponds to one line; circles correspond to H and He\,I and squares to ions.} 
\end{figure}  

\begin{figure}
\includegraphics[angle=-90,width=0.7\textwidth,bb=30 60 550 780,clip]{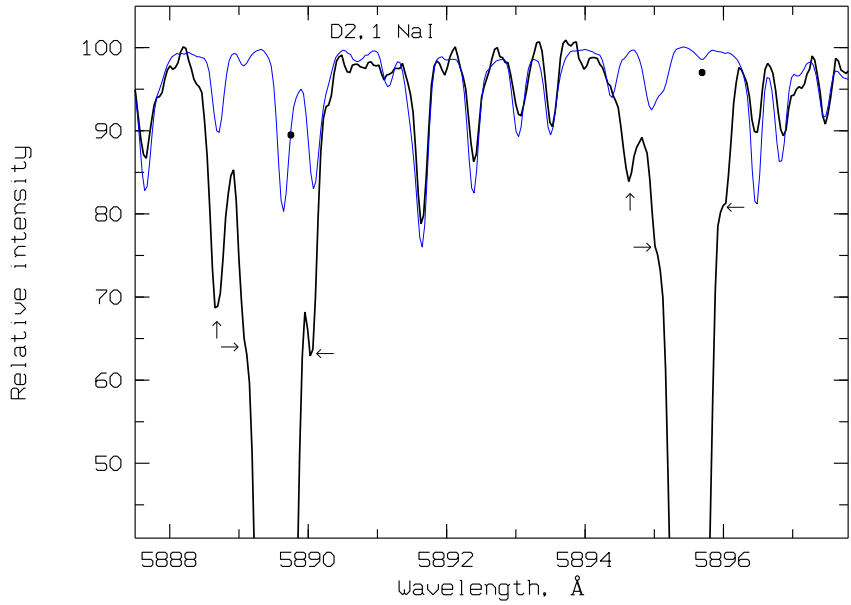}  
\caption{Section of the spectrum with interstellar sodium and atmospheric water--vapor absorption lines. 
         The blue curve is the spectrum of the comparison star HR\,4687. Apart from the weak interstellar 
         Na\,I\,(1) lines, marked with dots, all the other lines in the spectrum are telluric. 
         The bold curve is the spectrum of star No.\,6; the vertical arrows indicate the profile components 
         with Vr\,=\,$-50$\,km/s and the horizontal arrows those with Vr\,=\,$-29$ and 20\,km/s.} 
\end{figure}

\begin{figure}
\caption{Profiles of the interstellar D2\,Na\,I~5889.95\,\AA{} line in the spectra of Cyg~OB2 No.\,9 
         (black curve) and No.\,22 (red curve). For comparison, the dashed curve shows the lower part 
         of the profile for star No.\,2,  the narrowest in our sample. Telluric lines were excluded.}
\includegraphics[angle=0,width=0.45\textwidth,height=0.65\textwidth,bb=30 30 580 780,clip]{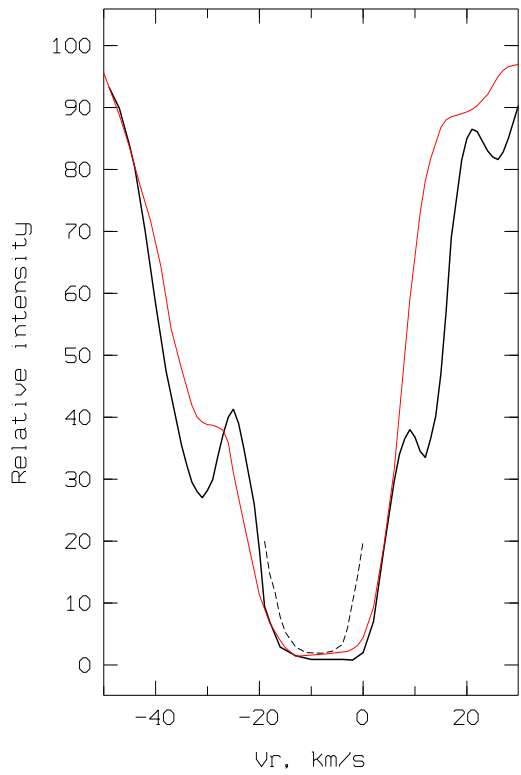}
\end{figure}

\begin{figure}
\hbox{
\includegraphics[angle=0,width=0.5\textwidth,bb=25 30 550 780,clip]{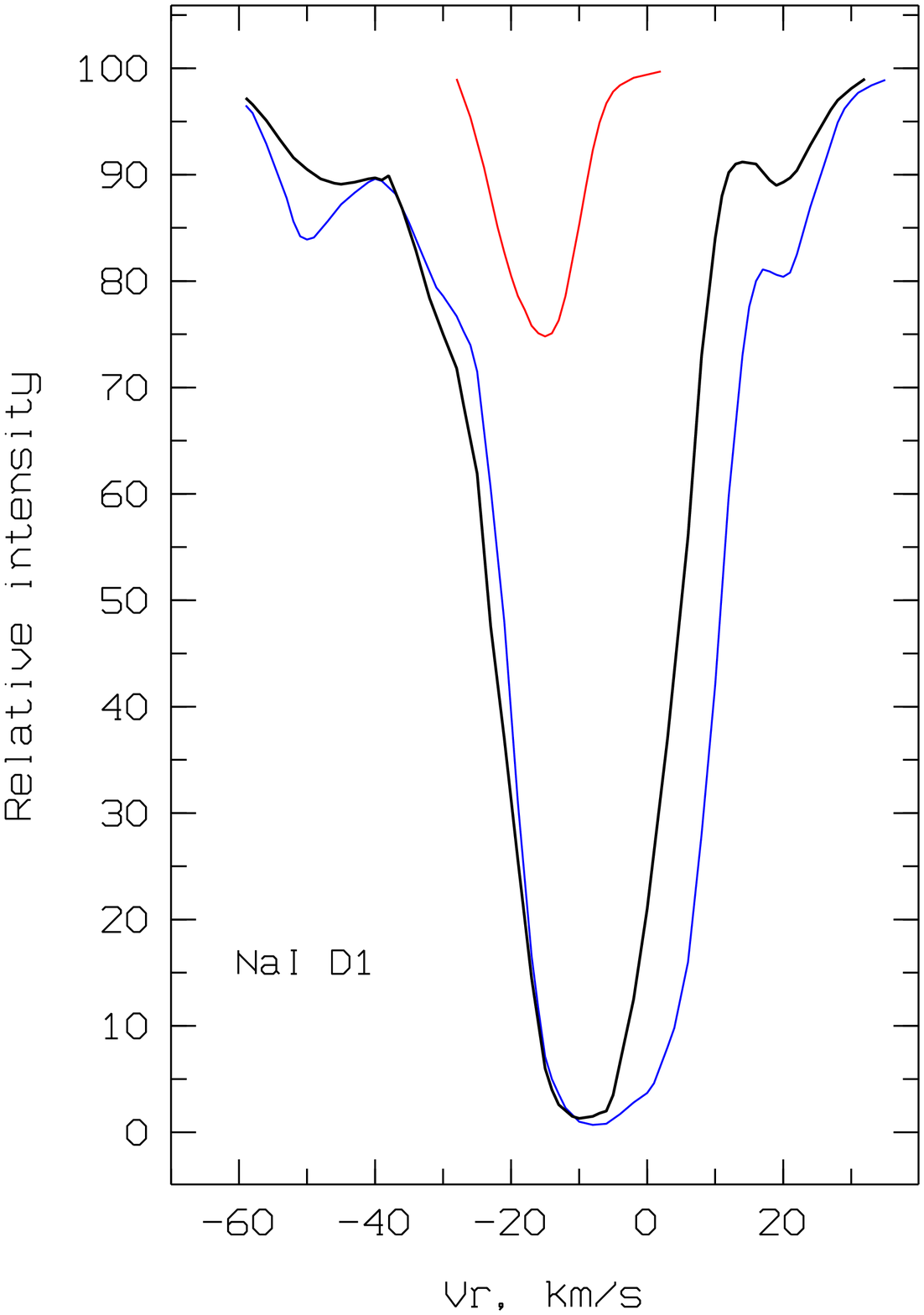} 
\includegraphics[angle=0,width=0.5\textwidth,bb=25 30 550 780,clip]{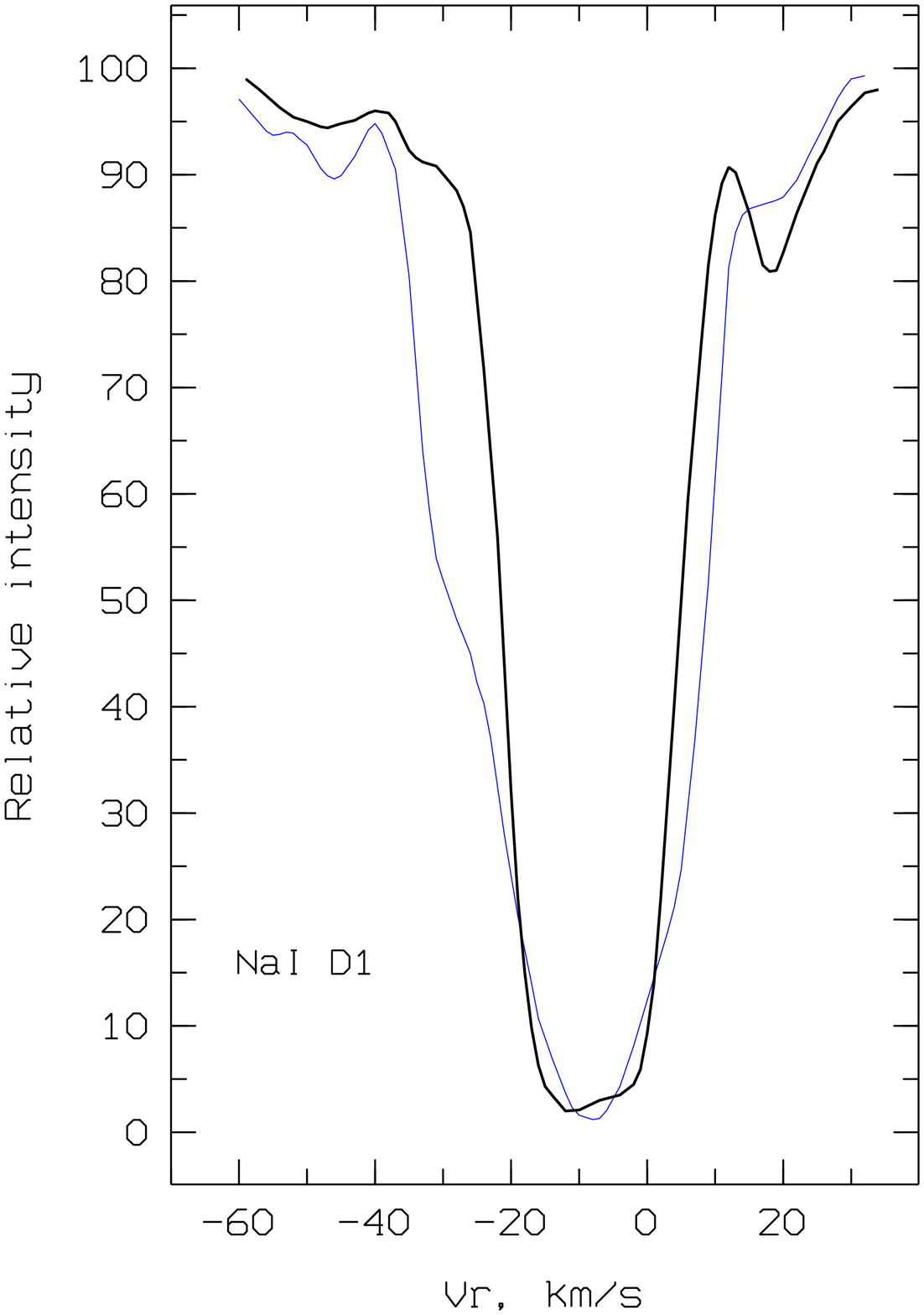} 
}
\caption{Profiles of the interstellar D1\,NaI~(1) 5895.92\,\AA{} line in the spectra of Cyg~OB2 stars No.\,16\, 
        (left, black curve),  No.\,6 (left, blue curve), No.\,4 (right, black curve), and No.\,21 (right, blue curve). 
        The left panel also displays the profile  of the interstellar D1 line in the spectrum of the foreground star 
        BD\,+41$^{\rm o}$\,3814 (red curve in the upper part of the panel).  Telluric lines were excluded.} 
\end{figure}  

\clearpage

\begin{figure}
\includegraphics[angle=-90,width=0.95\textwidth,bb=30 40 550 780,clip]{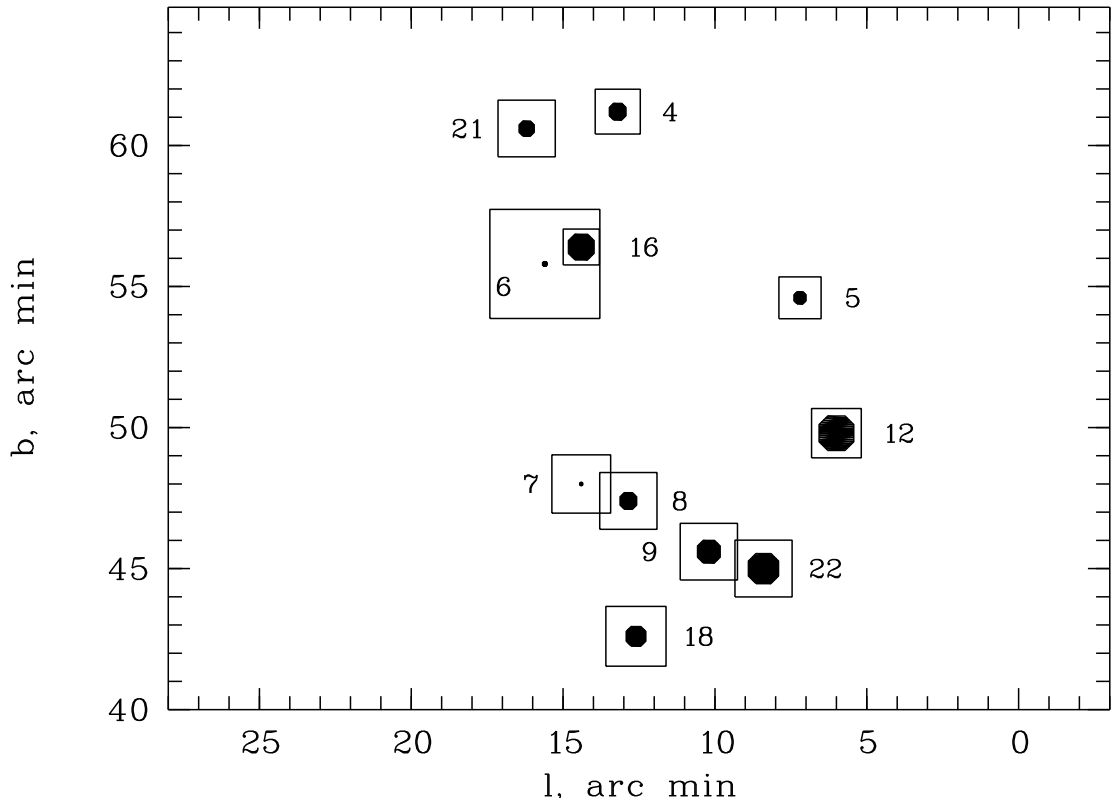}
\caption{Profile broadening of the main line components toward the blue (filled circles) and toward the red (squares) for stars in
         the central part of the Cyg~OB2 association, compared to the profile of star No.\,2 (see text for explanation). The size of the
         symbols is proportional to the broadening. The star numbers are from Schulte~[4]. 
         Galactic coordinates are plotted: $l-80^{\rm o}$ along the horizontal and  $b$ along the vertical axes.} 
\end{figure}

\begin{figure}
\vbox{
\includegraphics[angle=0,width=0.45\textwidth,bb=30 40 550 780,clip]{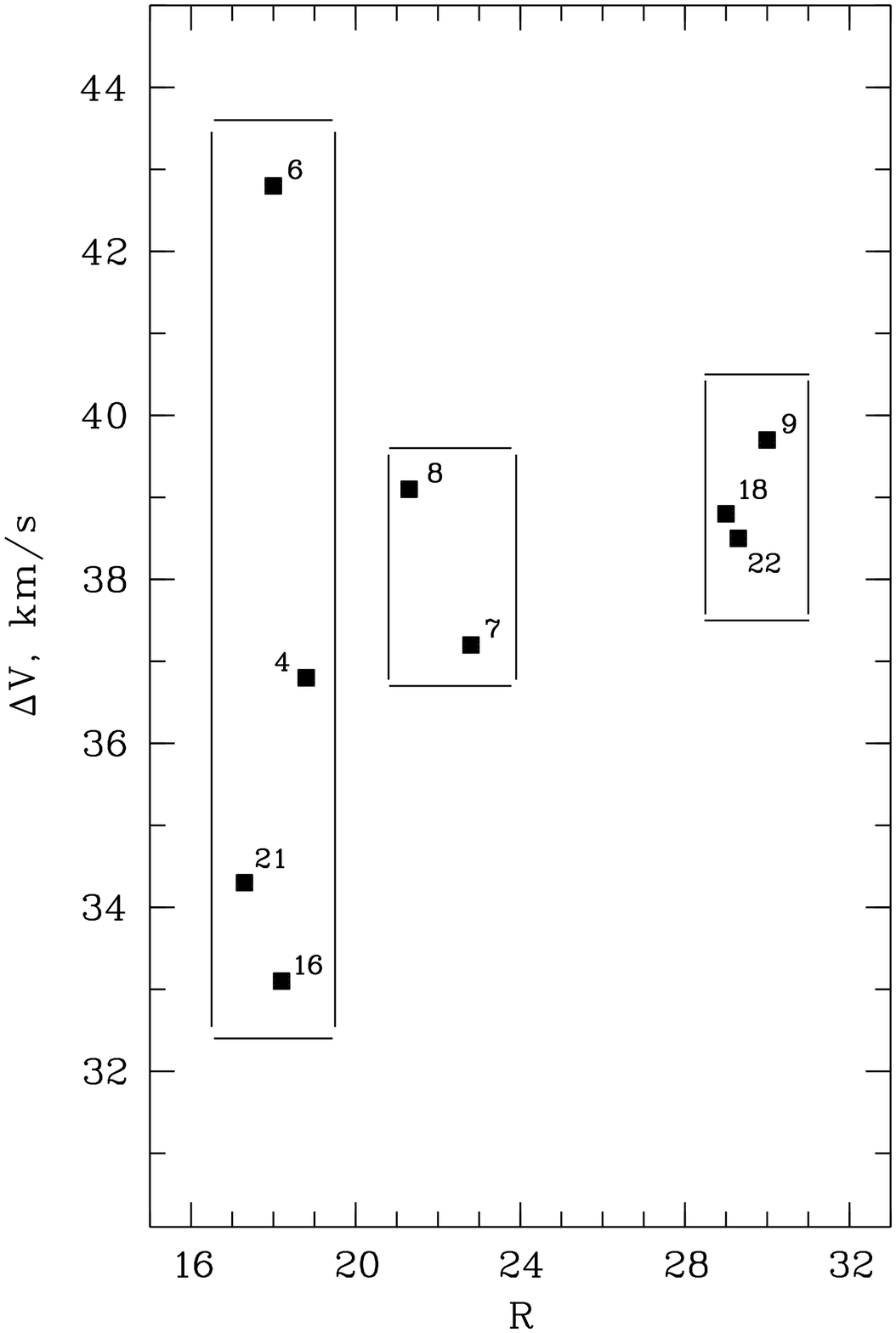}
\bigskip
\includegraphics[angle=0,width=0.45\textwidth,bb=30 40 550 780,clip]{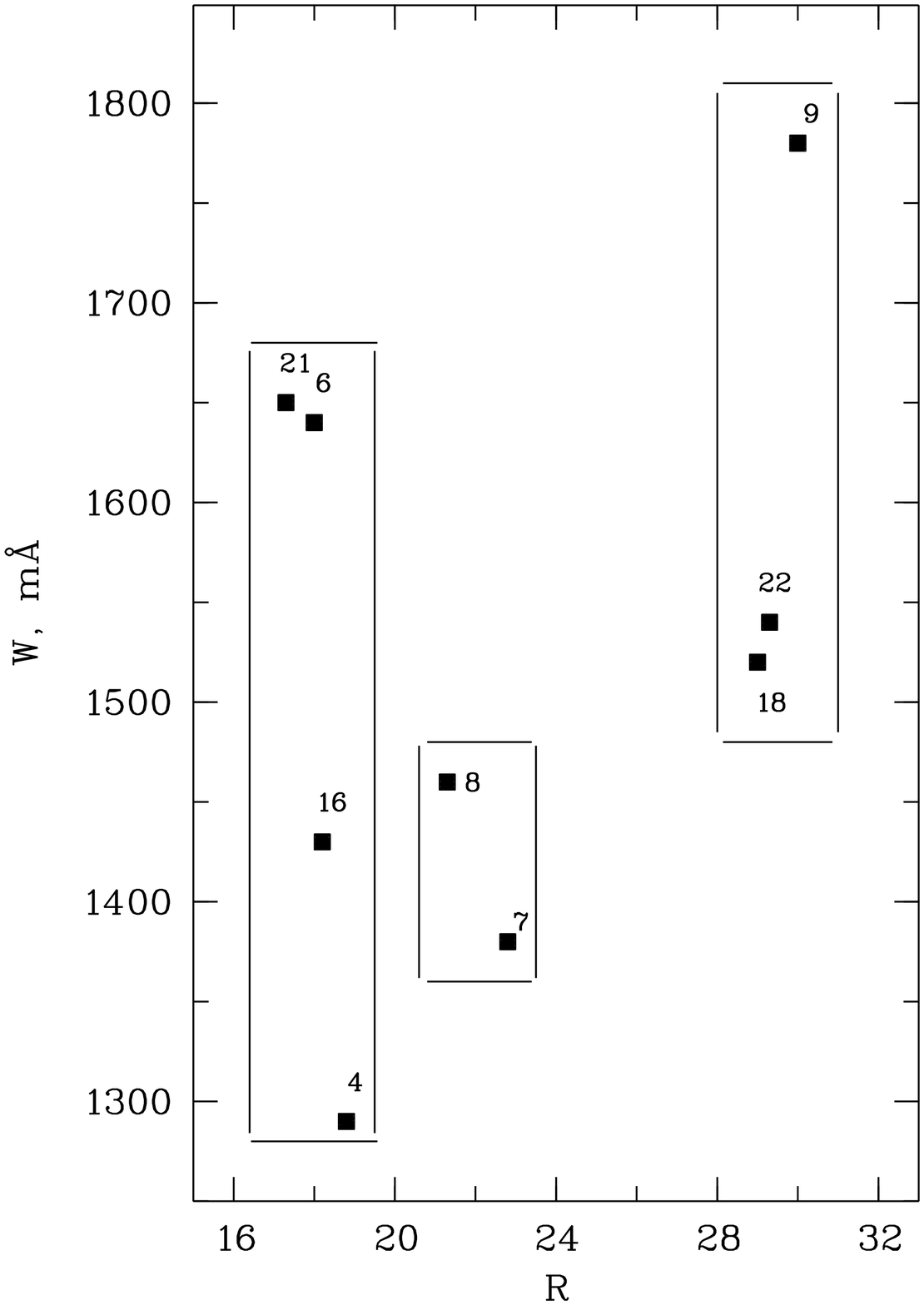} 
}
\caption{Left: total width $\Delta$V and  right: sum of equivalent widths W for lines of the Na\,I\,(1) doublet 
       at the r\,=\,10 level versus the  DIB. As in Figs.\,8 and 9, only stars in the association’s central  
       part are plotted; the rectangles delineate depth of the 5797\,\AA{} compact groups of these stars.}
\end{figure}

\begin{figure}
\includegraphics[angle=-90,width=0.85\textwidth,bb=30 40 550 780,clip]{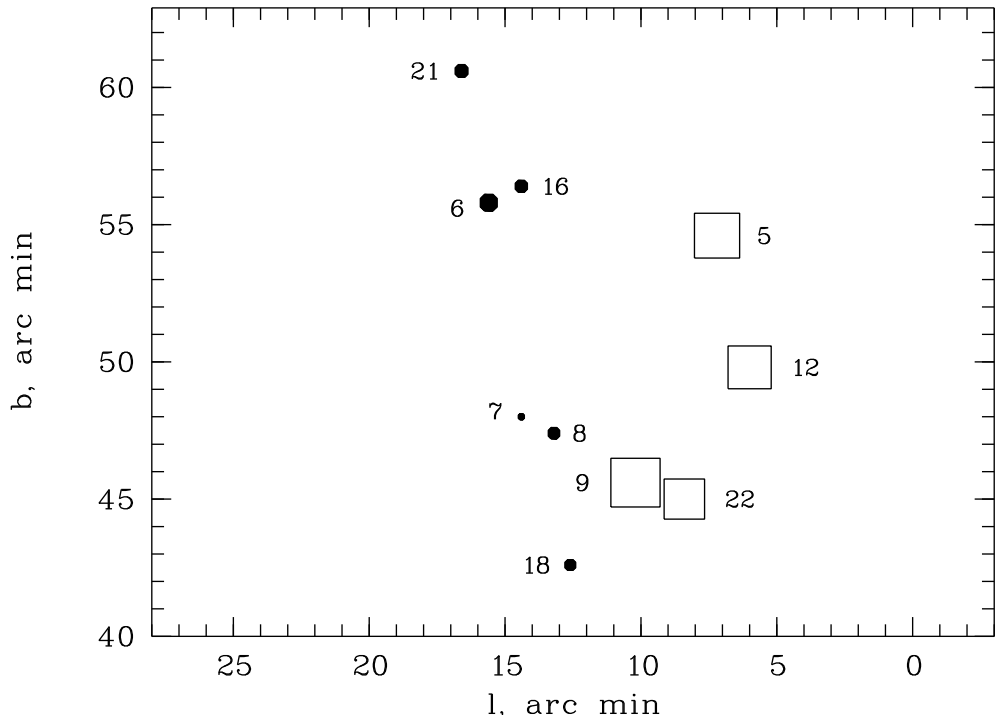}
\caption{The blue side components of the D2~Na\,I line for stars in the same region of the Cyg~OB2 association
        as in Fig.\,8. The filled circles and squares show the components with Vr$\approx -50\pm2$\,km/s 
           and Vr$\approx -32\pm2$\,km/s, respectively. The symbol sizes are proportional to the component depths.}
\end{figure}

\begin{figure}
\hbox{
\includegraphics[angle=0,width=0.5\textwidth,bb=30 40 550 780,clip]{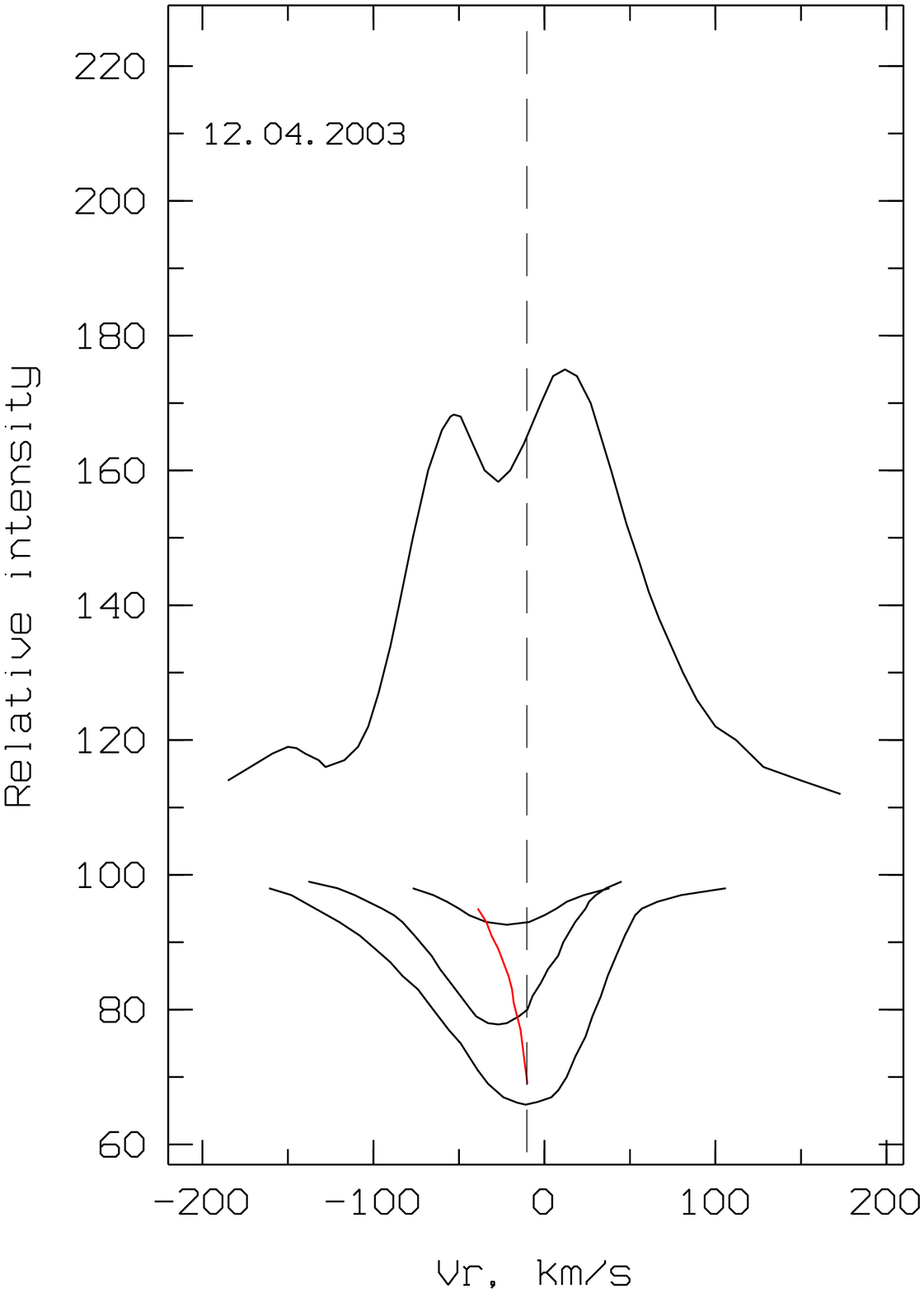} 
\bigskip
\includegraphics[angle=0,width=0.5\textwidth,bb=30 40 550 780,clip]{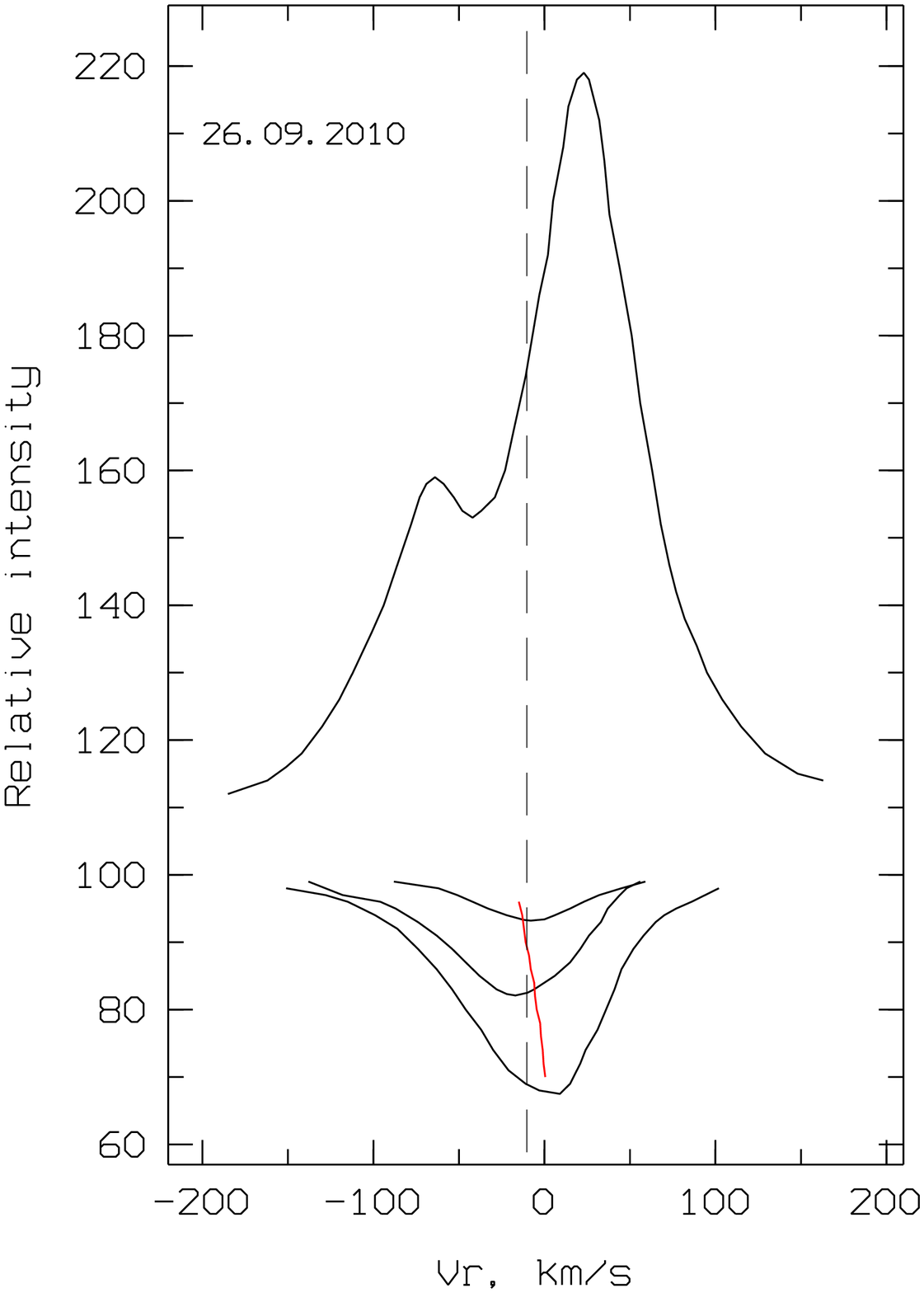} 
}
\caption{From top to bottom:  profiles of the H$\alpha$, Si\,III~5740\,\AA{}, Si\,II 6347\,\AA{}, 
         and He\,I~5876\,\AA{} lines,  in the spectra of star No.\,12 for two different dates. The red curves 
         bisect the HeI~5876\,\AA{} line profiles. The dashed vertical line is the systemic velocity.} 
\end{figure}

\begin{figure}
\hbox{
\hspace{-2cm}
\includegraphics[angle=-90,width=0.6\textwidth,bb=30 40 560 780,clip]{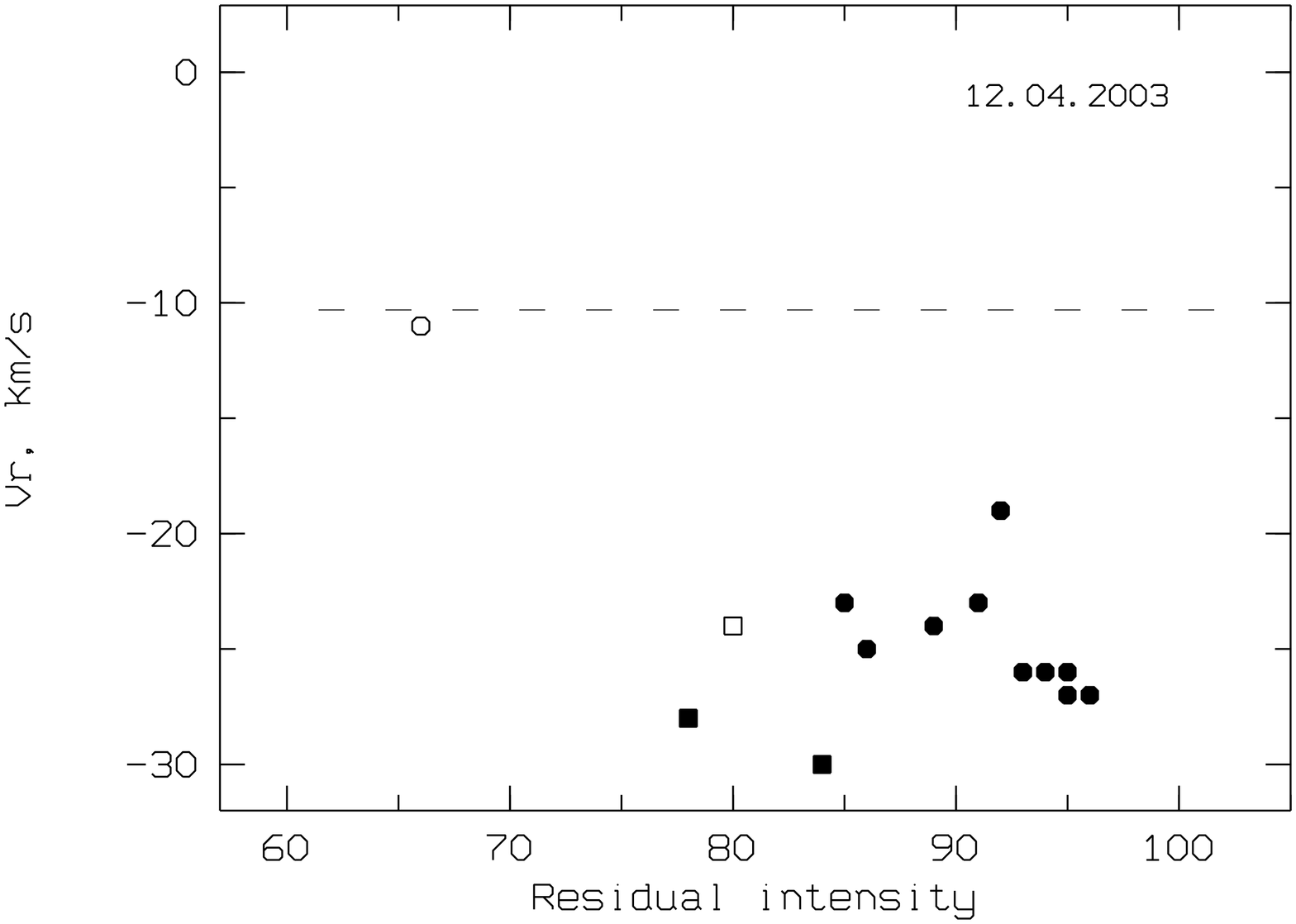} 
\bigskip
\includegraphics[angle=-90,width=0.6\textwidth,bb=30 40 560 780,clip]{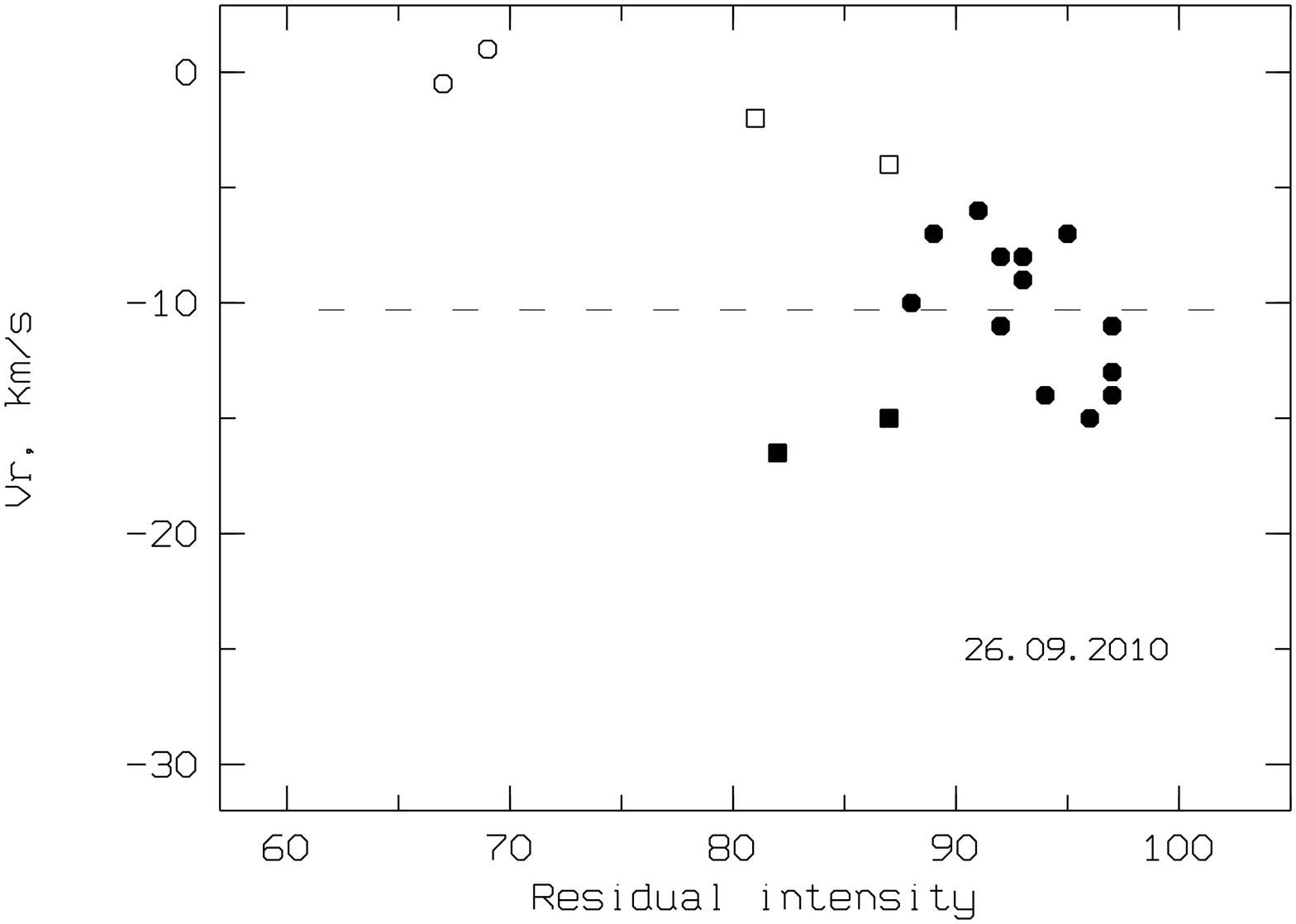} 
}
\caption{Radial velocity plotted against residual intensity for two spectra of star No.\,12, taken in 
        2003~(left) and  2010~(right). Each symbol represents one line. The filled circles are lines of 
        N\,II, S\,II, etc.; open circles He\,I lines; ﬁlled squares Si\,II lines; and open squares C\,II lines. 
        The dashed horizontal line corresponds  to the systemic velocity.} 
\end{figure}

\begin{figure}
\includegraphics[angle=0,width=0.5\textwidth,bb=30 40 550 780,clip]{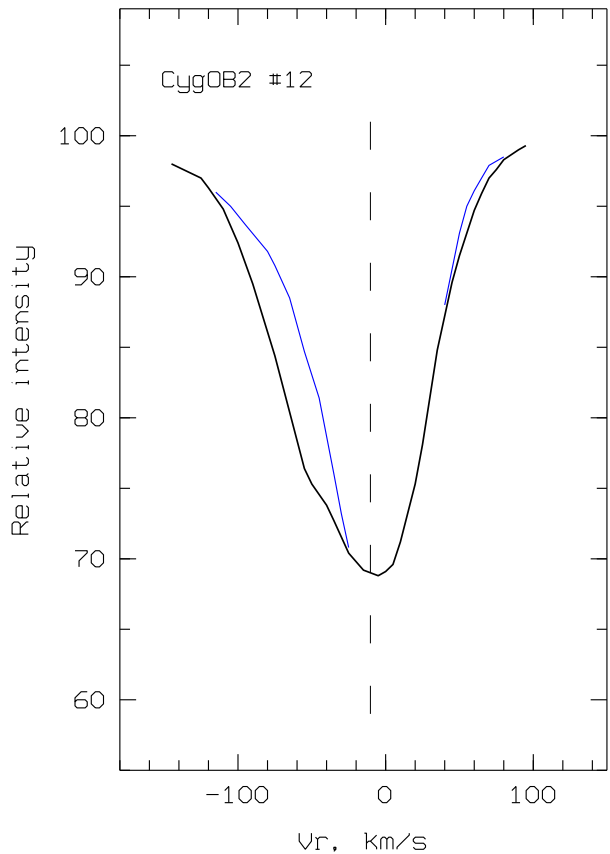} 
\caption{Profiles of the He\,I~5876\,\AA{} star No.\,12 taken on December~8, 2006 (bold curve) and
         September~15, 2011 (thin curve; the spectrum has been shifted horizontally so that 
         the two line cores coincide).The vertical dashed line corresponds to the systemic velocity.} 
\end{figure}

\begin{figure}
\includegraphics[angle=0,width=0.5\textwidth,bb=30 40 550 780,clip]{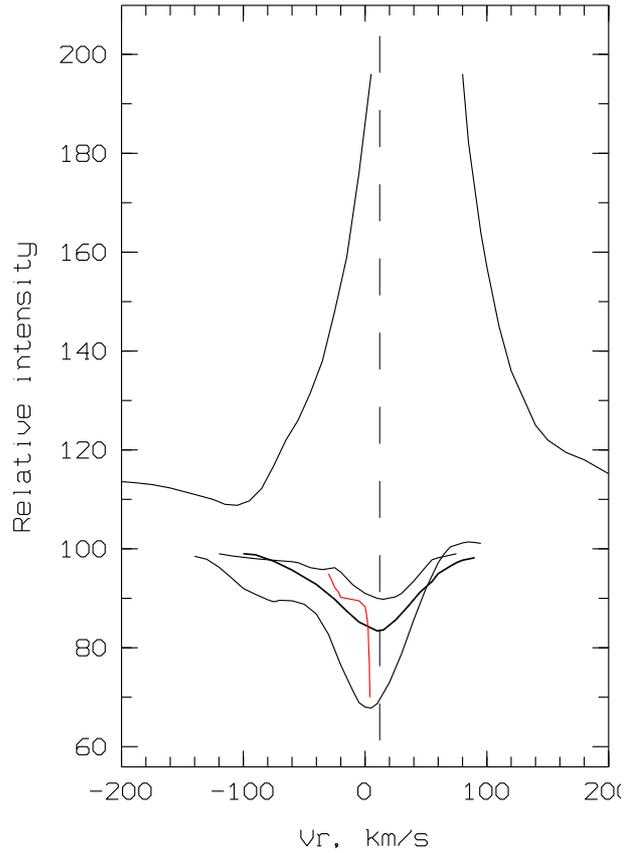} 
\caption{Same as Fig.~11 for the star HD\,80077. The  order of the Si\,II~6347\,\AA{} and Si\,III~5740\,\AA{} 
        profiles is  opposite compared to Fig.~11: HD\,80077 is hotter than Cyg~OB2 No.~12.}  
\end{figure}

\begin{figure}
\includegraphics[angle=0,height=0.8\textwidth,bb=30 40 550 780,clip]{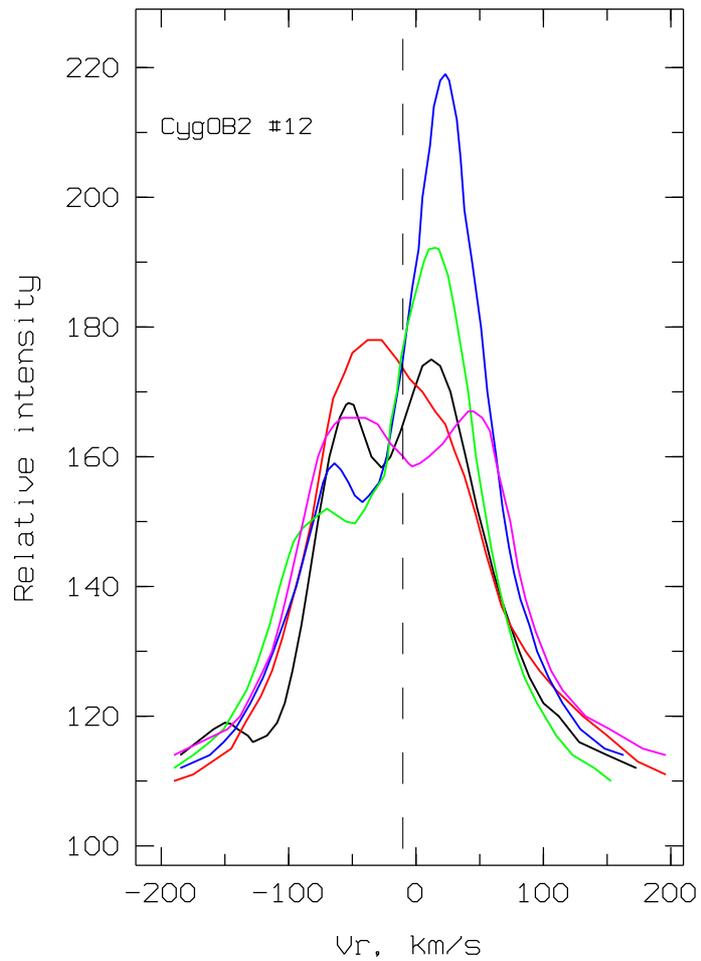} 
\caption{Variations of the H$\alpha$ profile in the spectrum of star No.\,12. Telluric lines have been excluded. The
         vertical dashed line indicates the systemic velocity.}
\end{figure}

\begin{figure}
\includegraphics[angle=0,width=0.7\textwidth,bb=30 40 550 780,clip]{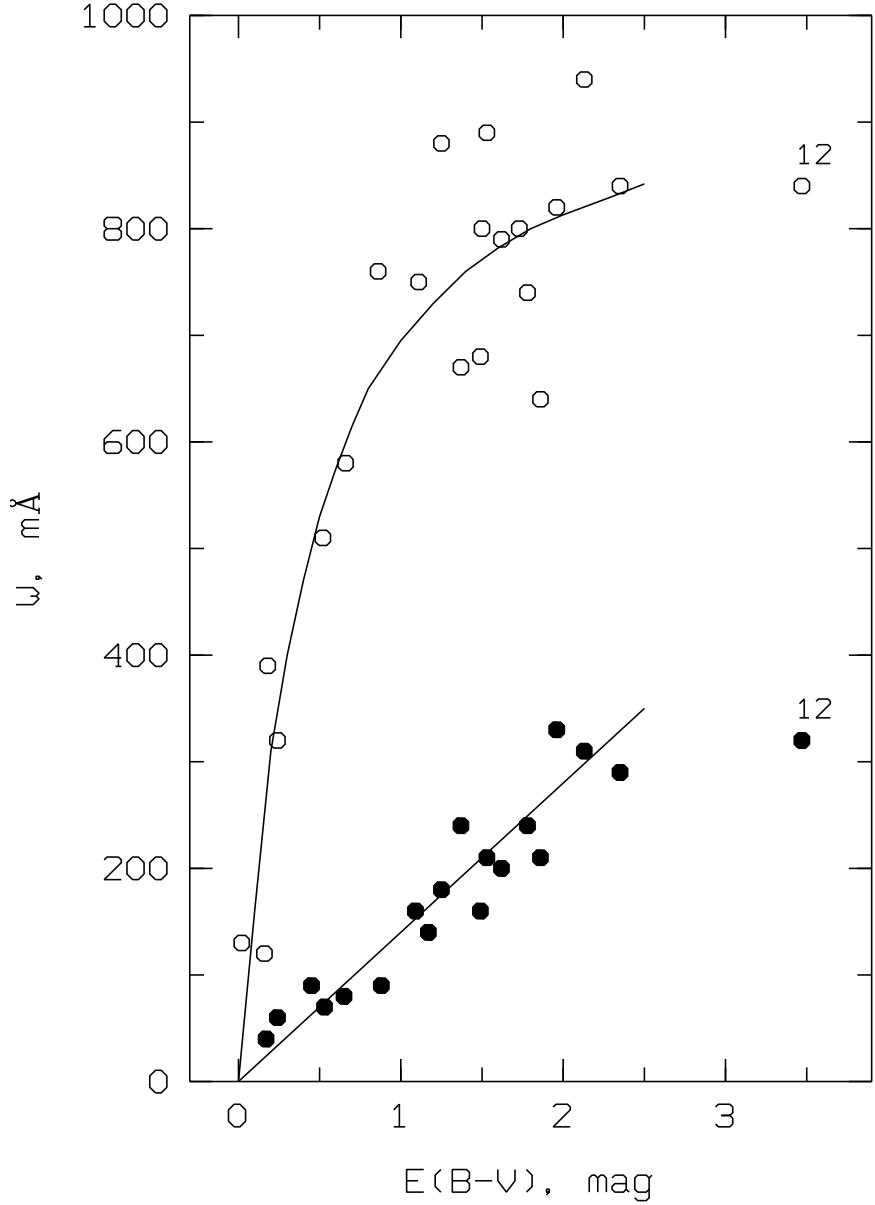} 
\caption{Dependence of the equivalent width W on the color excess E(B--V) for members of Cyg~OB2 (E\,$> 1.2^m$) 
         and foreground stars.  The open circles show results for the D2~NaI line and the filled circles results for the DIB~5797\,\AA{}.} 
\end{figure}

\end{document}